\documentclass[twocolumn]{jpsj3}
\usepackage{graphicx}
\usepackage{color}
\usepackage{bm}

\title{%
Bulk--Boundary Correspondence in a Non-Hermitian Quantum Spin-Hall Insulator
}

\author{%
Chihiro Ishii and Yositake Takane
}

\inst{%
Graduate School of Advanced Science and Engineering,\\
Hiroshima University, Higashihiroshima, Hiroshima 739-8530, Japan
}

\recdate{ \hspace{50mm} }

\abst{%
We focus on a scenario of non-Hermitian bulk--boundary correspondence
that uses a topological invariant defined in a bulk geometry
under a modified periodic boundary condition.
Although this has succeeded in describing the topological nature of
various one-dimensional non-Hermitian systems,
its application to two-dimensional systems has been limited to
a non-Hermitian Chern insulator.
Here, we adapt the scenario to a non-Hermitian quantum spin-Hall insulator
to extend its applicability.
We show that it properly describes the bulk--boundary correspondence
in the non-Hermitian quantum spin-Hall insulator.
A phase diagram derived from the bulk--boundary correspondence
is shown to be consistent with spectra of the system
under an open boundary condition.
}

\begin{document}
\maketitle

\section{Introduction}

Bulk--boundary correspondence is a key feature of topological systems,
including quantum Hall insulators,~\cite{thouless,kohmoto,hatsugai}
Chern insulators,~\cite{haldane,qi} quantum spin-Hall insulators
(i.e., two-dimensional topological insulators),~\cite{kane,bernevig}
and topological superconductors.~\cite{kitaev1,ivanov,fu2,sato1,sato2}
The original scenario of bulk--boundary correspondence~\cite{hatsugai,ryu2}
employs two geometries: a bulk geometry
under a periodic boundary condition (pbc)
and a boundary geometry under an open boundary condition (obc).
The scenario guarantees that a topological invariant defined
in the bulk geometry predicts the presence or absence of
topological boundary states in the boundary geometry.

Considerable attention has been paid to non-Hermitian versions of
various topological systems:~\cite{rudner,hu,esaki,ghosh}
one-dimensional topological insulators,~\cite{zhu,t_lee,xiong,yao1,alvarez,
yokomizo1,imura1,koch,he,yokomizo2,imura2,ccliu,kunst1,song,herviou,yuce1}
Chern insulators,~\cite{kunst1,song,herviou,yuce1,yao2,kawabata1,
borgnia,takane1,takane2,masuda}
quantum spin-Hall insulators,~\cite{kawabata2}
topological semimetals,~\cite{xu2,zyuzin,okugawa,papaj,yokomizo3}
one-dimensional superconductors,~\cite{wang2,yuce2,zeng,klett,menke,li1,
kawabata3,okuma1,zhao,sakaguchi,sayyad}
correlated electron systems,~\cite{ashida,yoshida1,yoshida2,e_lee,yoshida3}
and others.~\cite{yuce3,gong1,zhou1,li2,zhou2,mochizuki1,wu1,li3,bessyo,
yuce4,malzard,mochizuki2,leykam,wu2,kondo,kawasaki,yokomizo4,mostafavi}
Non-Hermitian topological insulators and superconductors are classified
in an exhaustive manner.~\cite{kawabata4}.
The bulk--boundary correspondence is also extensively studied
in the non-Hermitian regime (see Sect.~III of Ref.~\citen{bergholtz}),
leading to an observation that the correspondence is broken~\cite{t_lee,xiong}
in the presence of a non-Hermitian skin effect.
The non-Hermitian skin effect~\cite{yao1,longhi1,c_lee,kunst2,okuma2,zhang,
yi,longhi2} indicates the fact that eigenfunctions of
a non-Hermitian system under the obc tend to be localized
near a boundary of the system.
Note that this effect vanishes under the pbc.
Since the topological invariant calculated in the bulk geometry under the pbc
does not reflect the non-Hermitian skin effect,
it cannot predict the presence or absence of
topological boundary states in the boundary geometry.

Several scenarios~\cite{yao1,yokomizo1,kunst1,borgnia}
without using the bulk geometry have been
proposed to overcome this difficulty.
The scenario proposed by Yao and Wang~\cite{yao1}
for one-dimensional topological insulators is expected to
have wide applicability (see Sect.~V of Ref.~\citen{kawabata4}).
In this scenario, a topological invariant is calculated
in a generalized Brillouin zone 
that describes the bulk spectrum of a non-Hermitian system under the obc.
Its extension to two-dimensional systems seems not easy
since a tractable method of determining the generalized Brillouin zone
has been available only for one-dimensional systems.~\cite{yokomizo1}

Let us focus on another scenario employing
the bulk and boundary geometries,~\cite{imura1,imura2}
where the bulk geometry is defined under a modified periodic boundary condition
(mpbc) [see Eqs.~(\ref{eq:BC^R_mpbc}) and (\ref{eq:BC^L_mpbc})].
The mpbc enables us to take into account the non-Hermitian skin effect
in a closed system without a boundary.
This scenario is applicable even in the presence of
the non-Hermitian skin effect
and has been successfully applied to
one-dimensional topological insulators~\cite{imura1,imura2},
topological superconductors,~\cite{sakaguchi}
and Chern insulators.~\cite{takane1,takane2}
To show its wide applicability, we should provide more examples,
particularly those in two-dimensional systems.

In this paper,
we demonstrate that the scenario~\cite{imura1,imura2}
correctly describes the bulk--boundary correspondence
in a non-Hermitian quantum spin-Hall insulator.
We use a revised scenario that is successfully applied to
a non-Hermitian Chern insulator.~\cite{takane2}
Our model for a non-Hermitian quantum spin-Hall insulator
with gain/loss-type non-Hermiticity shows a topologically trivial phase,
a nontrivial phase with helical edge states, and a gapless phase.
We show that a phase diagram derived from the bulk--boundary correspondence
correctly specifies the phase realized in the boundary geometry.

In the next section, we present a tight-binding Hamiltonian
for the non-Hermitian quantum spin-Hall insulator.
In Sect.~3, we describe plane-wave-like basis states
in the bulk geometry under the mpbc,
which are used to define a topological invariant in Sect.~4.
In Sect.~4, after introducing the $\mathbb{Z}_{2}$ invariant
characterizing the trivial and nontrivial phases,
we execute the bulk--boundary correspondence
and obtain a phase diagram in the boundary geometry.
In Sect.~5, we justify the resulting phase diagram by comparing it
with spectra of the system in the boundary geometry.
The last section is devoted to a summary.

\section{Model and Symmetries}

We consider a tight-binding model for
a non-Hermitian quantum spin-Hall insulator
on a square lattice with the lattice constant $a$,
where each lattice site has orbital and spin degrees of freedom described
by $\eta = 1, 2$ and $\sigma = \uparrow, \downarrow$, respectively.
We use $m$ and $n$ to specify the location of each site
in the $x$- and $y$-directions, respectively.
Let us introduce four component basis vectors
$|m,n \rangle$ and $\langle m,n|$ for the $(m,n)$th site:
\begin{align}
  |m,n \rangle
  & = \Bigl[ |m,n \rangle_{1\uparrow} \hspace{1.2mm}
             |m,n \rangle_{2\uparrow} \hspace{1.2mm}
             |m,n \rangle_{1\downarrow} \hspace{1.2mm}
             |m,n \rangle_{2\downarrow}
      \Bigr] ,
  \\
  \langle m,n|
  & = \left[ \begin{array}{c}
               {}_{1\uparrow}\langle m,n| \\
               {}_{2\uparrow}\langle m,n| \\
               {}_{1\downarrow}\langle m,n| \\
               {}_{2\downarrow}\langle m,n|
             \end{array}
      \right] .
\end{align}
Here, $|m,n \rangle_{\eta\sigma}$ and ${}_{\eta\sigma}\langle m,n|$
are respectively the column and row vectors
consisting of $4N_{\rm site}$ components,
where $N_{\rm site}$ is the number of sites in the system.
In the $4N_{\rm site}$ components of
$|m,n \rangle_{\eta\sigma}$ and ${}_{\eta\sigma}\langle m,n|$,
only one component corresponding to the $(m,n)$th site
with the given $\eta$ and $\sigma$ is $1$ and the others are $0$.
They satisfy ${}_{\eta\sigma}\langle m,n| = {}^{t}|m,n \rangle_{\eta\sigma}$
and
\begin{align}
   {}_{\eta\sigma}\langle m,n|m',n' \rangle_{\eta'\sigma'}
    = \delta_{m,m'}\delta_{n,n'}
      \delta_{\eta,\eta'}\delta_{\sigma,\sigma'} .
\end{align}
The Hamiltonian is given by
$H = H_{\rm d}+H_{x}+H_{y}$ with
\begin{align}
   H_{\rm d}
 & = \sum_{m,n} |m,n \rangle \mathcal{H}_{d} \langle m,n| ,
         \\
   H_{x}
 & = \sum_{m,n} |m+1,n \rangle \mathcal{H}_{x} \langle m,n| + {\rm h.c.} ,
        \\
   H_{y}
 & = \sum_{m,n} |m,n+1 \rangle \mathcal{H}_{y} \langle m,n| + {\rm h.c.} ,
\end{align}
where
\begin{align}
   \mathcal{H}_{d}
 & = \left[ 
       \begin{array}{cc}
         M\tau_{z}+i\gamma\left(\tau_{x}+\tau_{y}\right) &
         \alpha(1-i)i\gamma\tau_{x} \\
         \alpha(1+i)i\gamma\tau_{x} &
         M\tau_{z}-i\gamma\left(\tau_{x}-\tau_{y}\right)
       \end{array}
     \right] ,
       \\
   \mathcal{H}_{x}
 & = \left[ 
       \begin{array}{cc}
         \frac{i}{2}\tau_{x} - \frac{1}{2}\tau_{z} &
         \frac{i}{2}\alpha\tau_{x} \\
         \frac{i}{2}\alpha\tau_{x} &
         -\frac{i}{2}\tau_{x} - \frac{1}{2}\tau_{z}
       \end{array}
     \right] ,
       \\
   \mathcal{H}_{y}
 & = \left[ 
       \begin{array}{cc}
         \frac{i}{2}\tau_{y} - \frac{1}{2}\tau_{z} &
         \frac{1}{2}\alpha\tau_{x} \\
         -\frac{1}{2}\alpha\tau_{x} &
         \frac{i}{2}\tau_{y} - \frac{1}{2}\tau_{z}
       \end{array}
     \right] .
\end{align}
Here, $M$, $\gamma$, and $\alpha$ are real dimensionless parameters,
and $\tau_{q}$ is the $q$-component of Pauli matrices ($q \in x,y$, and $z$).
The non-Hermiticity of $H$ is characterized by $\gamma$ in $\mathcal{H}_{d}$.
We consider the case of $M \ge 0$ with $\gamma \ge 0$ throughout this paper.
In the Hermitian limit of $\gamma = 0$, the model is essentially equivalent to that given in Ref.~\citen{bernevig}
and becomes topologically nontrivial when $0 < M < 2$.

The Hamiltonian $H$ exhibits three symmetries.
The first one is the time reversal symmetry,
which protects the quantum spin-Hall insulator phase.
This is expressed as
\begin{align}
  {\cal T}H^{*}{\cal T}^{-1} = H
\end{align}
with
\begin{align}
  \cal{T} = \left[ \begin{array}{cc}
                      0_{2 \times 2} & -\tau_{0} \\
                      \tau_{0} & 0_{2 \times 2}
                    \end{array}
             \right] ,
\end{align}
where $\tau_{0}$ is the $2 \times 2$ identity matrix
and $\cal{T}$ acts on the $4 \times 4$ space corresponding to
the orbital and spin degrees of freedom.
From this symmetry, we can show for $|\Psi^{R}\rangle$ satisfying
$H|\Psi^{R}\rangle = E |\Psi^{R}\rangle$ that
$H\left({\cal T}|\Psi^{R}\rangle\right)^{*}
= E^{*}\left({\cal T}|\Psi^{R}\rangle\right)^{*}$.
That is, an eigenstate of $H$ with an eigenvalue $E$ is paired with
its counterpart with the eigenvalue $E^{*}$.
The second symmetry is expressed as
\begin{align}
      \label{eq:U-symmetry}
  U H^{*} U^{-1} = - H ,
\end{align}
where $U$ acting on the $4 \times 4$ space is
\begin{align}
  U = \frac{1}{\sqrt{1+\alpha^{2}}}
      \left[ \begin{array}{cc}
               \tau_{x} & \alpha\tau_{x} \\
               \alpha\tau_{x} & -\tau_{x}
             \end{array}
      \right] .
\end{align}
From this symmetry, we can show for $|\Psi^{R}\rangle$ satisfying
$H|\Psi^{R}\rangle = E |\Psi^{R}\rangle$ that
$H \left(U|\Psi^{R}\rangle\right)^{*}
= -E^{*}\left(U|\Psi^{R}\rangle\right)^{*}$.
That is, an eigenstate of $H$ with an eigenvalue $E$
is paired with its counterpart with the eigenvalue $-E^{*}$.

The third one manifests itself if the geometry of the system
is inversion symmetric with respect to $m$ and $n$.
Hereafter,
we assume that the system is square shaped with $N \times N$ sites.
The third one is expressed as
\begin{align}
      \label{eq:V-symmetry}
  PV \,{}^{t}\!H \,{}^{t}\!(PV) = H ,
\end{align}
where $V$ acting on the $4 \times 4$ space is
\begin{align}
  V = \left[ \begin{array}{cc}
               0_{2 \times 2} & -\tau_{z} \\
               \tau_{z} & 0_{2 \times 2}
             \end{array}
      \right]
\end{align}
and $P$ is given by
\begin{align}
  P = \sum_{m,n}|N+1-m,N+1-n \rangle \langle m,n| .
\end{align}
From this symmetry, we can show for $|\Psi^{R}\rangle$ satisfying
$H|\Psi^{R}\rangle = E |\Psi^{R}\rangle$ that
${}^{t}\!\left(PV|\Psi^{R}\rangle\right)H
= {}^{t}\!\left(PV|\Psi^{R}\rangle\right) E$.
That is, $E$ is doubly degenerate if $|\Psi^{R}\rangle$ and
${}^{t}\!\left(PV|\Psi^{R}\rangle\right)$ are orthogonal.

The first and second symmetries require that if an eigenstate
with an eigenvalue $E$ is present,
this eigenstate and its three counterparts form a quartet:
four eigenstates with $E$, $E^{*}$, $-E^{*}$, and $-E$ are related
by the two symmetries.

\section{Bulk Geometry}

We consider the system of $N \times N$ sites in the bulk geometry.
The right and left eigenstates of $H$ are respectively written as
\begin{align}
      \label{eq:right-eigenvec}
 &  |\Psi^{R}\rangle
    = \sum_{m,n} |m,n \rangle \cdot |\psi^{R}(m,n)\rangle ,
     \\
 &  \langle\Psi^{L}|
    = \sum_{m,n} \langle\psi^{L}(m,n)| \cdot \langle m,n| ,
\end{align}
where
\begin{align}
  |\psi^{R}(m,n)\rangle
  & = \left[ \begin{array}{c}
               \psi_{1\uparrow}^{R}(m,n) \\
               \psi_{2\uparrow}^{R}(m,n) \\
               \psi_{1\downarrow}^{R}(m,n) \\
               \psi_{2\downarrow}^{R}(m,n)
             \end{array}
      \right] ,
     \\
  \langle\psi^{L}(m,n)|
  & = \Bigl[ \psi_{1\uparrow}^{L}(m,n) \hspace{1mm}
             \psi_{2\uparrow}^{L}(m,n) \hspace{1mm}
             \psi_{1\downarrow}^{L}(m,n) \hspace{1mm}
             \psi_{2\downarrow}^{L}(m,n)
      \Bigr] .
\end{align}
To define a topological invariant, we introduce plane-wave-like right and left
basis states by setting
\begin{align}
    \label{eq:psi^R}
  |\psi^{R}(m,n)\rangle
  & = \varphi^{R}(m,n) |\psi^{R}\rangle ,
      \\
    \label{eq:psi^L}
  \langle\psi^{L}(m,n)|
  & = \langle \psi^{L}| \varphi^{L}(m,n) ,
\end{align}
where $|\psi^{R}\rangle$ and $\langle\psi^{L}|$ are respectively
four component column and row vectors independent of $m$ and $n$, and
\begin{align}
  \varphi^{R}(m,n) & = \beta_{x}^{m}\beta_{y}^{n} ,
    \\
  \varphi^{L}(m,n) & = \beta_{x}^{-m}\beta_{y}^{-n} .
\end{align}
Under the mpbc with a positive real constant $b$:
\begin{align}
      \label{eq:BC^R_mpbc}
 \varphi^{R}(m+N,n) = \varphi^{R}(m,n+N) = b^{N}\varphi^{R}(m,n) ,
     \\
      \label{eq:BC^L_mpbc}
 \varphi^{L}(m+N,n) = \varphi^{L}(m,n+N) = b^{-N}\varphi^{L}(m,n) ,
\end{align}
we find
\begin{align}
 & \beta_{x} = be^{ik_{x}a},
   \hspace{3mm} \beta_{y} = be^{ik_{y}a} ,
\end{align}
where $k_{i} = 2n_{i}\pi/N$ and $n_{i} = 0,1,2,\dots, N-1$ ($i = x, y$).
The resulting basis states are the right and left eigenstates of
$H_{\rm mpbc} = H + \Delta H_{x} + \Delta H_{y}$, where $\Delta H_{x}$ and
$\Delta H_{y}$ are the boundary terms in accordance with the mpbc:
\begin{align}
   \Delta H_{x}
 & = \sum_{n}
     \left( |1,n \rangle b^{-N}\mathcal{H}_{x} \langle N,n|
            + |N,n \rangle b^{N}\mathcal{H}_{x}^{\dagger} \langle 1,n|
     \right) ,
        \\
   \Delta H_{y}
 & = \sum_{m}
     \left( |m,1 \rangle b^{-N}\mathcal{H}_{y} \langle m,N|
            + |m,N \rangle b^{N}\mathcal{H}_{y}^{\dagger} \langle m,1|
     \right) .
\end{align}
The time reversal symmetry is preserved in $H_{\rm mpbc}$.
The eigenvalue equations $H_{\rm mpbc}|\Psi^{R}\rangle = E|\Psi^{R}\rangle$
and $\langle \Psi^{L}|H_{\rm mpbc} = \langle \Psi^{L}|E$ are respectively
reduced to
\begin{align}
  h|\psi^{R}\rangle = E|\psi^{R}\rangle ,
    \\
  \langle \psi^{L}|h = \langle \psi^{L}|E ,
\end{align}
where 
\begin{align}
  h = \left[ \begin{array}{cc}
               \eta_{x}\tau_{x}+\eta_{y}\tau_{y}+\eta_{z}\tau_{z} &
               \alpha \left(\eta_{x}-i\eta_{y}\right)\tau_{x} \\
               \alpha \left(\eta_{x}+i\eta_{y}\right)\tau_{x} &
               -\eta_{x}\tau_{x}+\eta_{y}\tau_{y}+\eta_{z}\tau_{z}
             \end{array}
      \right]
\end{align}
with
\begin{align}
     \label{eq:def-eta_x}
 & \eta_{x}
   = \frac{1}{2i}\left(\beta_{x}-\beta_{x}^{-1}\right)+i\gamma ,
      \\
     \label{eq:def-eta_y}
 & \eta_{y}
   = \frac{1}{2i}\left(\beta_{y}-\beta_{y}^{-1}\right)+i\gamma ,
      \\
     \label{eq:def-eta_z}
 & \eta_{z}
   = M - \frac{1}{2}\left(\beta_{x}+\beta_{x}^{-1}\right)
       - \frac{1}{2}\left(\beta_{y}+\beta_{y}^{-1}\right) .
\end{align}
For later convenience, we rewrite $\eta_{x}$, $\eta_{y}$, and $\eta_{z}$ as
\begin{align}
     \label{eq:def-eta_x-mod}
 & \eta_{x}
   = b_{+}\sin(k_{x}a) +i\left[ \gamma - b_{-}\cos(k_{x}a) \right] ,
      \\
     \label{eq:def-eta_y-mod}
 & \eta_{y}
  = b_{+}\sin(k_{y}a) +i\left[ \gamma - b_{-}\cos(k_{y}a) \right] ,
      \\
 & \eta_{z}
   = M - b_{+}\left[\cos(k_{x}a)+\cos(k_{y}a)\right]
      \nonumber \\
 & \hspace{20mm}
       -ib_{-}\left[ \sin(k_{x}a) + \sin(k_{y}a) \right] .
\end{align}
where
\begin{align}
    b_{\pm} = \frac{1}{2}\left(b \pm b^{-1}\right) .
\end{align}
From the time reversal symmetry, we can show that
the reduced Hamiltonian $h(k_{x},k_{y},b)$ satisfies
\begin{align}
     \label{eq:TRS-k}
  {\cal T} h(k_{x},k_{y},b)^{*} {\cal T}^{-1} = h(-k_{x},-k_{y},b) .
\end{align}

Solving the reduced eigenvalue equation, we find that the energy of
an eigenstate characterized by $\mib{k}=(k_{x},k_{y})$ and $b$
is given by $E = \pm \epsilon(\mib{k})$ with
\begin{align}
    \label{eq:def-epsilon}
 \epsilon(\mib{k})
 = \sqrt{(1+\alpha^2)\left(\eta_{x}^{2}+\eta_{y}^{2}\right)+\eta_{z}^{2}} ,
\end{align}
where $\Re\{\epsilon\} \ge 0$.
A set of $\epsilon(\mib{k})$ ($-\epsilon(\mib{k})$) with all allowed $\mib{k}$
is referred to as the conduction (valence) band.
The right and left eigenstates of $H_{\rm mpbc}$
with the eigenvalue $\pm\epsilon(\mib{k})$ are respectively expressed as
\begin{align}
    |\Psi_{\zeta\pm}^{R}(\mib{k})\rangle
  & = \frac{1}{N} \sum_{m,n}
      \beta_{x}^{m}\beta_{y}^{n} |m,n \rangle
      \cdot |\psi_{\zeta\pm}^{R}(\mib{k})\rangle ,
     \\
    \langle \Psi_{\zeta\pm}^{L}(\mib{k})|
  & = \frac{1}{N} \sum_{m,n}
      \langle\psi_{\zeta\pm}^{L}(\mib{k})| \cdot \langle m,n|
      \beta_{x}^{-m}\beta_{y}^{-n} ,
\end{align}
where $\zeta = 1$, $2$ is used to specify two branches
in the conduction and valence bands, and
\begin{align}
       \label{eq:psi_k1_R}
  |\psi_{1 \pm}^{R}(\mib{k})\rangle
  & = c_{\pm}(\mib{k})
      \left[ \begin{array}{c}
               \eta_{z}\pm\epsilon \\
               \eta_{x}+i\eta_{y} \\
               0 \\
               \alpha \left(\eta_{x}+i\eta_{y}\right)
             \end{array}
      \right] ,
      \\
       \label{eq:psi_k2_R}
  |\psi_{2 \pm}^{R}(\mib{k})\rangle
  & = c_{\pm}(\mib{k})
      \left[ \begin{array}{c}
               0 \\
               \alpha \left(\eta_{x}-i\eta_{y}\right) \\
               \eta_{z}\pm\epsilon \\
               -\eta_{x}+i\eta_{y}
             \end{array}
      \right] ,
      \\
       \label{eq:psi_k1_L}
  ^{t}\! \langle\psi_{1 \pm}^{L}(\mib{k})|
  & = c_{\pm}(\mib{k})
      \left[ \begin{array}{c}
               \eta_{z}\pm\epsilon \\
               \eta_{x}-i\eta_{y} \\
               0 \\
               \alpha \left(\eta_{x}-i\eta_{y}\right)
             \end{array}
      \right] ,
      \\
       \label{eq:psi_k2_L}
  ^{t}\! \langle\psi_{2 \pm}^{L}(\mib{k})|
  & = c_{\pm}(\mib{k})
      \left[ \begin{array}{c}
               0 \\
               \alpha \left(\eta_{x}+i\eta_{y}\right) \\
               \eta_{z}\pm\epsilon \\
               -\eta_{x}-i\eta_{y}
             \end{array}
       \right]
\end{align}
with
\begin{align}
 c_{\pm}(\mib{k})
 = \frac{1}
        {\sqrt{(1+\alpha^{2})\left(\eta_{x}^{2}+\eta_{y}^{2}\right)
                +(\eta_{z}\pm\epsilon)^{2}}} .
\end{align}
The eigenvectors satisfy
\begin{align}
 \langle\psi_{\zeta\pm}^{L}(\mib{k})|\psi_{\zeta'\pm}^{R}(\mib{k})\rangle
  = \delta_{\zeta,\zeta'} ,
 \hspace{2mm}
 \langle\psi_{\zeta\pm}^{L}(\mib{k})|\psi_{\zeta'\mp}^{R}(\mib{k})\rangle = 0 .
\end{align}
We easily find that
\begin{align}
  \langle \Psi_{\zeta\pm}^{L}(\mib{k})|\Psi_{\zeta'\pm}^{R}(\mib{k'})\rangle
  & = \delta_{\mib{k},\mib{k'}}\delta_{\zeta,\zeta'} ,
  \\
  \langle \Psi_{\zeta\pm}^{L}(\mib{k})|\Psi_{\zeta'\mp}^{R}(\mib{k'})\rangle
  & = 0 .
\end{align}
That is, $\{|\Psi_{\zeta\pm}^{R}(\mib{k})\rangle\}$
and $\{\langle \Psi_{\zeta\pm}^{L}(\mib{k})|\}$
constitute a biorthogonal set of basis functions for a given $b$.

Let $\Theta = {\cal T}K$ be the time reversal operator
with $K$ denoting complex conjugation.
Equation~(\ref{eq:TRS-k}) requires that
\begin{align}
     \label{eq:result1-TRS}
  \Theta |\psi_{2-}^{R}(\mib{k})\rangle
  & = e^{-i\chi(\mib{k})} |\psi_{1-}^{R}(-\mib{k})\rangle ,
  \\
     \label{eq:result2-TRS}
  \Theta |\psi_{1-}^{R}(\mib{k})\rangle
  & = -e^{-i\chi(-\mib{k})}|\psi_{2-}^{R}(-\mib{k})\rangle ,
\end{align}
where
\begin{align}
   e^{-i\chi(\mib{k})} = -\frac{c_{-}(\mib{k})^{*}}{c_{-}(-\mib{k})} .
\end{align}
Equations~(\ref{eq:result1-TRS}) and (\ref{eq:result2-TRS})
are used in Sect.~4.

\section{Bulk--Boundary Correspondence}

The plane-wave-like basis states in the bulk geometry form
the conduction and valence bands.
When the two bands are separated by a line gap
(i.e., ${\rm Re}\{\epsilon(\mib{k})\} \neq 0$ for an arbitrary $\mib{k}$),
we can define the $\mathbb{Z}_{2}$ invariant $\nu$ by using
$|\psi_{\zeta-}^{R}(\mib{k})\rangle$ and $\langle\psi_{\zeta-}^{L}(\mib{k})|$.
Here, we treat $b$ in the mpbc as a parameter characterizing
the system together with $\alpha$, $M$, and $\gamma$.
The matrix $w$, defined by
\begin{align}
  w(\mib{k})
  = \left[ \begin{array}{cc}
             \hspace{-1.5mm}\langle\psi_{1-}^{L}(-\mib{k})|
             \Theta|\psi_{1-}^{R}(\mib{k})\rangle &
             \hspace{-1.5mm}\langle\psi_{1-}^{L}(-\mib{k})|
             \Theta|\psi_{2-}^{R}(\mib{k})\rangle \hspace{-1.5mm} \\
             \hspace{-1.5mm}\langle\psi_{2-}^{L}(-\mib{k})|
             \Theta|\psi_{1-}^{R}(\mib{k})\rangle &
             \hspace{-1.5mm}\langle\psi_{2-}^{L}(-\mib{k})|
             \Theta|\psi_{2-}^{R}(\mib{k})\rangle \hspace{-1.5mm}
           \end{array}
    \right] ,
\end{align}
is reduced to
\begin{align}
  w(\mib{k})
  = \left[ \begin{array}{cc}
             0 & e^{-i\chi(\mib{k})} \\
             -e^{-i\chi(\mib{-k})} & 0
           \end{array}
    \right]
\end{align}
owing to Eqs.~(\ref{eq:result1-TRS}) and (\ref{eq:result2-TRS}).
Using $w_{12}(\mib{k}) = e^{-i\chi(\mib{k})}$
at the time reversal invariant momenta,
\begin{align}
   \Lambda_{1} = \left(0,0\right),
   \Lambda_{2} = \left(\frac{\pi}{a},0\right),
   \Lambda_{3} = \left(0,\frac{\pi}{a}\right),
   \Lambda_{4} = \left(\frac{\pi}{a},\frac{\pi}{a}\right) ,
\end{align}
we define the $\mathbb{Z}_{2}$ invariant $\nu$ as
\begin{align}
   \label{eq:def-nu}
  (-1)^{\nu}
  = \prod_{i =1,2,3,4}\frac{w_{12}(\Lambda_{i})}
                           {\sqrt{w_{12}(\Lambda_{i})^{2}}} .
\end{align}
The $\mathbb{Z}_{2}$ invariant takes $0$ or $1$
depending on $\alpha$, $M$, $\gamma$, and $b$.
In the Hermitian limit, Eq.~(\ref{eq:def-nu}) is reduced to
an ordinary expression of $\nu$.~\cite{kane}

Our system exhibits three phases: a topologically trivial phase
(i.e., an ordinary insulator phase) with $\nu = 0$,
a nontrivial phase (i.e., a quantum spin-Hall insulator phase) with $\nu = 1$,
and a gapless phase,
in which the line gap is closed and therefore $\nu$ cannot be defined.
We consider $\nu$ for the given $\alpha$ and $M$
in a parameter space spanned by $\gamma$ and $b$,
where these three phases are separated by lines on which the gap closes.
Let us find such gap-closing lines.~\cite{takane1}
Gap closing occurs when
$\Re \{\epsilon(\mib{k})\} = \Im \{\epsilon(\mib{k})\} = 0$.
Equation~(\ref{eq:def-epsilon}) indicates that
$\Im \{\epsilon(\mib{k})^{2}\} = 0$ at
$\mib{k} = \Lambda_{1}$, $\Lambda_{2}$, $\Lambda_{3}$, and $\Lambda_{4}$.
At $\mib{k}=\Lambda_{1}$, $\Re \{\epsilon(\mib{k})^{2}\} = 0$ when
\begin{align}
    \label{eq:gap-closing1}
  (M-2b_{+})^{2} - \zeta_{+}(\gamma-b_{-})^{2} = 0 .
\end{align}
Here and hereafter, we use $\zeta_{\pm}$ defined by
\begin{align}
   \zeta_{\pm} = 2(1 \pm \alpha^{2}) .
\end{align}
Equation~(\ref{eq:gap-closing1}) yields four gap-closing lines:
\begin{align}
      \label{eq:b_1}
  b_{\rm 1} & = \frac{M-\sqrt{\zeta_{+}}\gamma
                      + \sqrt{(M-\sqrt{\zeta_{+}}\gamma)^{2}-\zeta_{-}}}
                     {2-\sqrt{\zeta_{+}}} ,
         \\
      \label{eq:b_2}
  b_{\rm 2} & = \frac{M+\sqrt{\zeta_{+}}\gamma
                      + \sqrt{(M+\sqrt{\zeta_{+}}\gamma)^{2}-\zeta_{-}}}
                     {2+\sqrt{\zeta_{+}}} ,
         \\
      \label{eq:b_3}
  b_{\rm 3} & = \frac{M-\sqrt{\zeta_{+}}\gamma
                      - \sqrt{(M-\sqrt{\zeta_{+}}\gamma)^{2}-\zeta_{-}}}
                     {2-\sqrt{\zeta_{+}}} ,
         \\
      \label{eq:b_4}
  b_{\rm 4} & = \frac{M+\sqrt{\zeta_{+}}\gamma
                      - \sqrt{(M+\sqrt{\zeta_{+}}\gamma)^{2}-\zeta_{-}}}
                     {2+\sqrt{\zeta_{+}}} .
\end{align}
At $\mib{k} =\Lambda_{2}$ and $\Lambda_{3}$,
$\Re \{\epsilon(\mib{k})^{2}\} = 0$ when
\begin{align}
  M^{2}-\zeta_{+}\left(b_{-}^{2}+\gamma^{2}\right)= 0 ,
\end{align}
which yields two gap-closing lines:
\begin{align}
      \label{eq:b_5}
  b_{5} & =  \sqrt{\frac{M^{2}}{\zeta_{+}}-\gamma^{2}+1}
             + \sqrt{\frac{M^{2}}{\zeta_{+}}-\gamma^{2}} ,
         \\
      \label{eq:b_6}
  b_{6} & = \sqrt{\frac{M^{2}}{\zeta_{+}}-\gamma^{2}+1}
            - \sqrt{\frac{M^{2}}{\zeta_{+}}-\gamma^{2}} .
\end{align}
We ignore the case of $\mib{k} = \Lambda_{4}$
because this point has a gap in relevant situations.
Each $b_{i}$ ($i = 1, \dots, 6$) is irrelevant if it becomes a complex number.
In addition to $b_{i}$ ($i = 1, \dots, 6$), we should consider $b_{0}$
defined below if $0 \le \alpha < 1$ (i.e., $\zeta_{-} > 0$).
When
\begin{align}
  (M-\sqrt{\zeta_{+}}\gamma)^{2} < \zeta_{-} ,
\end{align}
$b_{1}$ and $b_{3}$ are expressed as $b_{0}e^{\pm ik_{0}a}$, where
\begin{align}
    \label{eq:b_0}
  b_{0} = \frac{2+\sqrt{\zeta_{+}}}{\sqrt{\zeta_{-}}} ,
\end{align}
and $k_{0}$ is defined by
\begin{align}
  \tan\left(k_{0}a\right)
   = \frac{\sqrt{\zeta_{-}-\left(M-\sqrt{\zeta_{+}}\gamma\right)^{2}}}
                    {M-\sqrt{\zeta_{+}}\gamma} .
\end{align}
Because $\Re \{\epsilon(\mib{k})^{2}\} = \Im \{\epsilon(\mib{k})^{2}\} = 0$
at $b = b_{0}$ if $\mib{k}=\pm (k_{0},k_{0})$,
$b_{0}$ is a gap-closing line when $0 \le \alpha < 1$.

We present distribution maps of $\nu$ in the $\gamma b$-plane
in the cases of $\alpha = 0.2$ and $1.2$.
Figures~1(a)--1(d) show the distribution maps in the case of
$\alpha = 0.2$ for $M = 1.2$, $2.1229$, $2.4$, and $3.5$,
where $b_{0}$ appears as a relevant gap-closing line.
Figures~2(a)--2(d) show the distribution maps in the case of
$\alpha = 1.2$ for $M = 1.5$, $2.2$, $2.6034$, and $4.0$,
where $b_{0}$ is absent.
In all of the figures, the gap-closing lines separate
the topologically trivial region ($\nu = 0$),
nontrivial region ($\nu = 1$), and gapless region.
In Figs.~1 and 2, any region in which $\nu$ is not specified
belongs to the gapless region.
\begin{figure}[btp]
\begin{tabular}{cc}
\begin{minipage}{0.5\hsize}
\begin{center}
\hspace{-10mm}
\includegraphics[height=3.8cm]{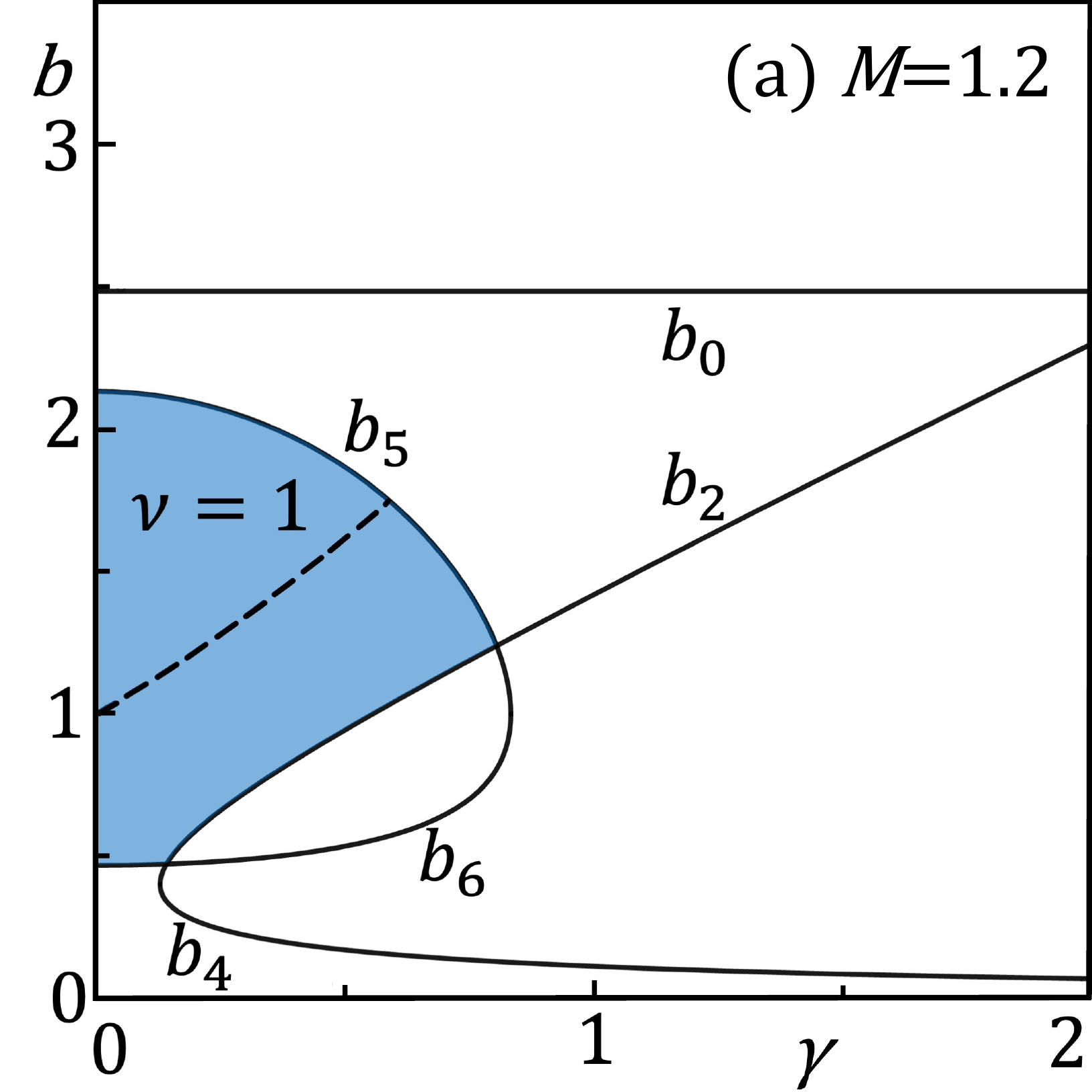}
\end{center}
\end{minipage}
\begin{minipage}{0.5\hsize}
\begin{center}
\hspace{-10mm}
\includegraphics[height=3.8cm]{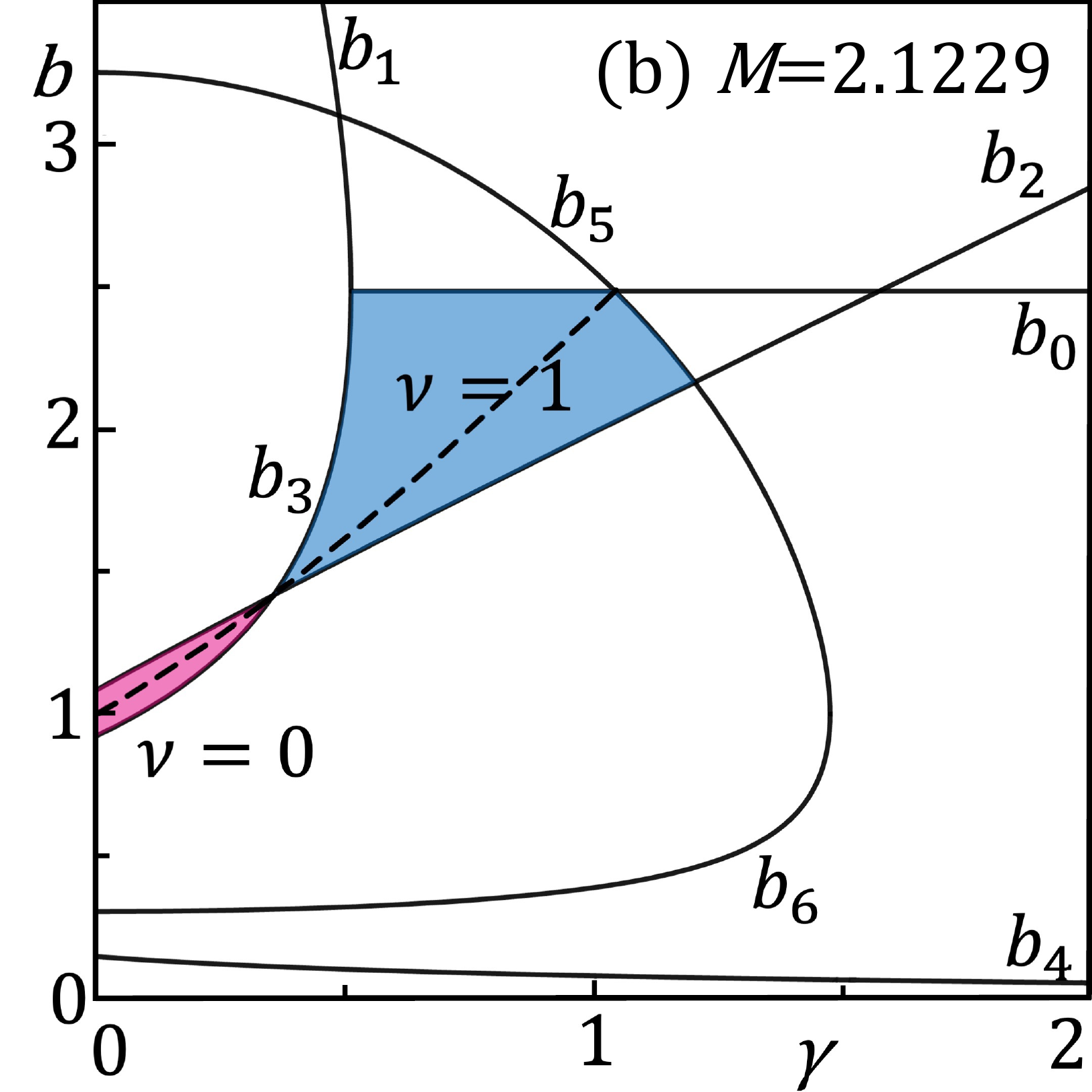}
\end{center}
\end{minipage}
\end{tabular}
\begin{tabular}{cc}
\begin{minipage}{0.5\hsize}
\begin{center}
\hspace{-10mm}
\includegraphics[height=3.8cm]{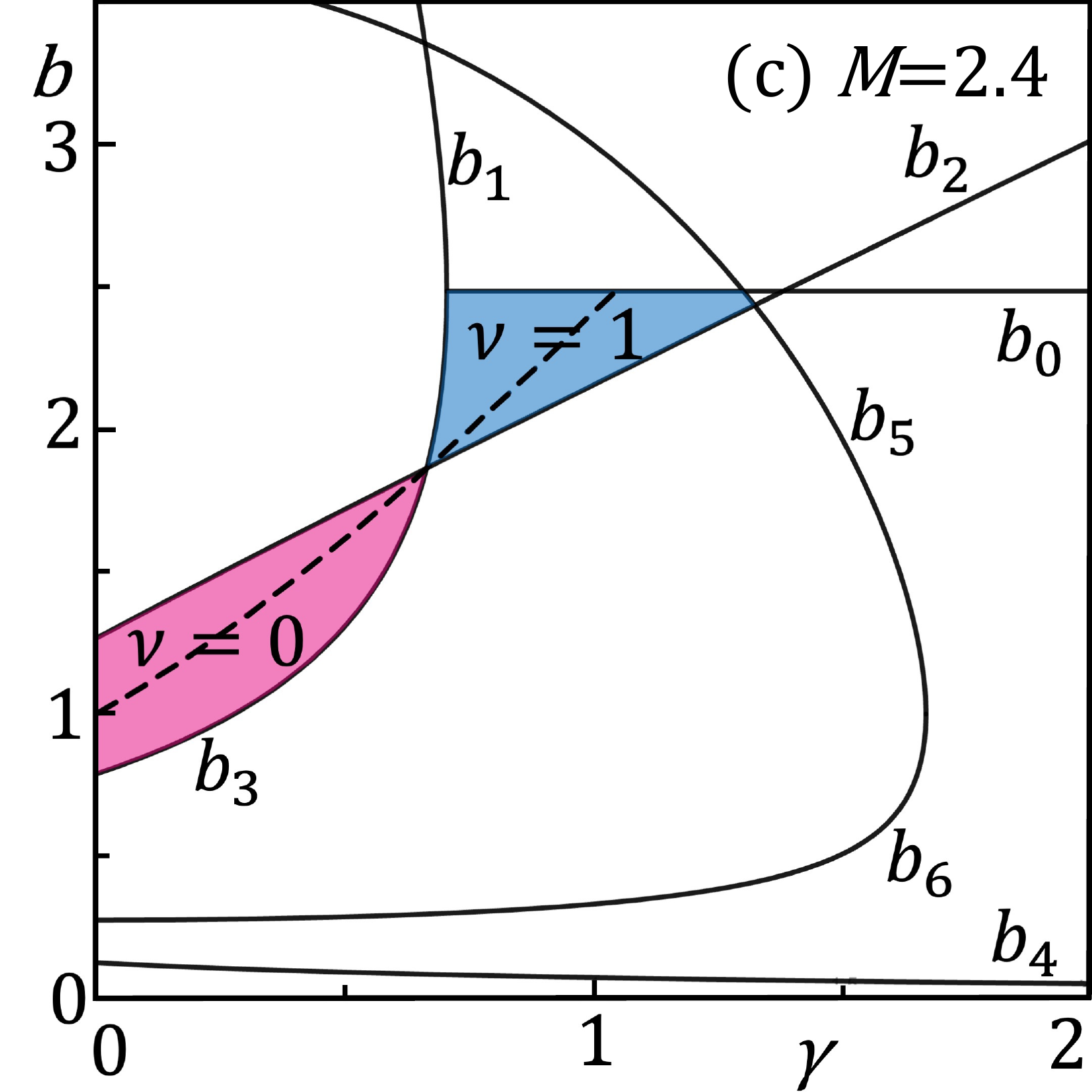}
\end{center}
\end{minipage}
\begin{minipage}{0.5\hsize}
\begin{center}
\hspace{-10mm}
\includegraphics[height=3.8cm]{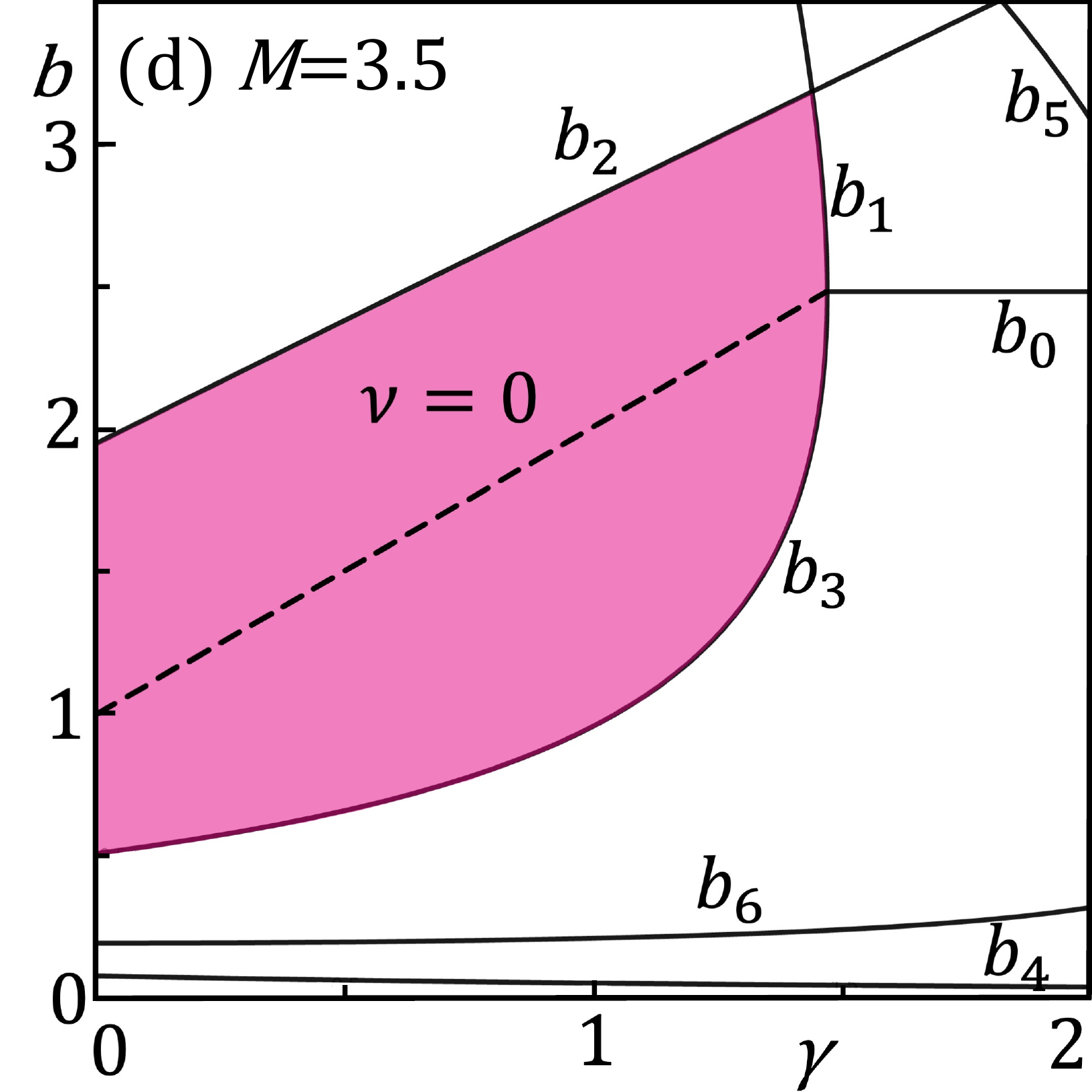}
\end{center}
\end{minipage}
\end{tabular}
\caption{
(Color online)
Distribution maps of $\nu$ in the $\gamma b$-plane for $\alpha = 0.2$
and $M =$ (a) $1.2$, (b) $2.1229$, (c) $2.4$, and (d) $3.5$.
Topologically trivial and nontrivial regions are respectively
designated as $\nu =0$ (mazenta) and $\nu =1$ (light blue),
outside of which is a gapless region.
In each figure, the dotted line represents a possible trajectory of
$b(\gamma)$.
}
\end{figure}

\begin{figure}[btp]
\begin{tabular}{cc}
\begin{minipage}{0.5\hsize}
\begin{center}
\hspace{-10mm}
\includegraphics[height=3.8cm]{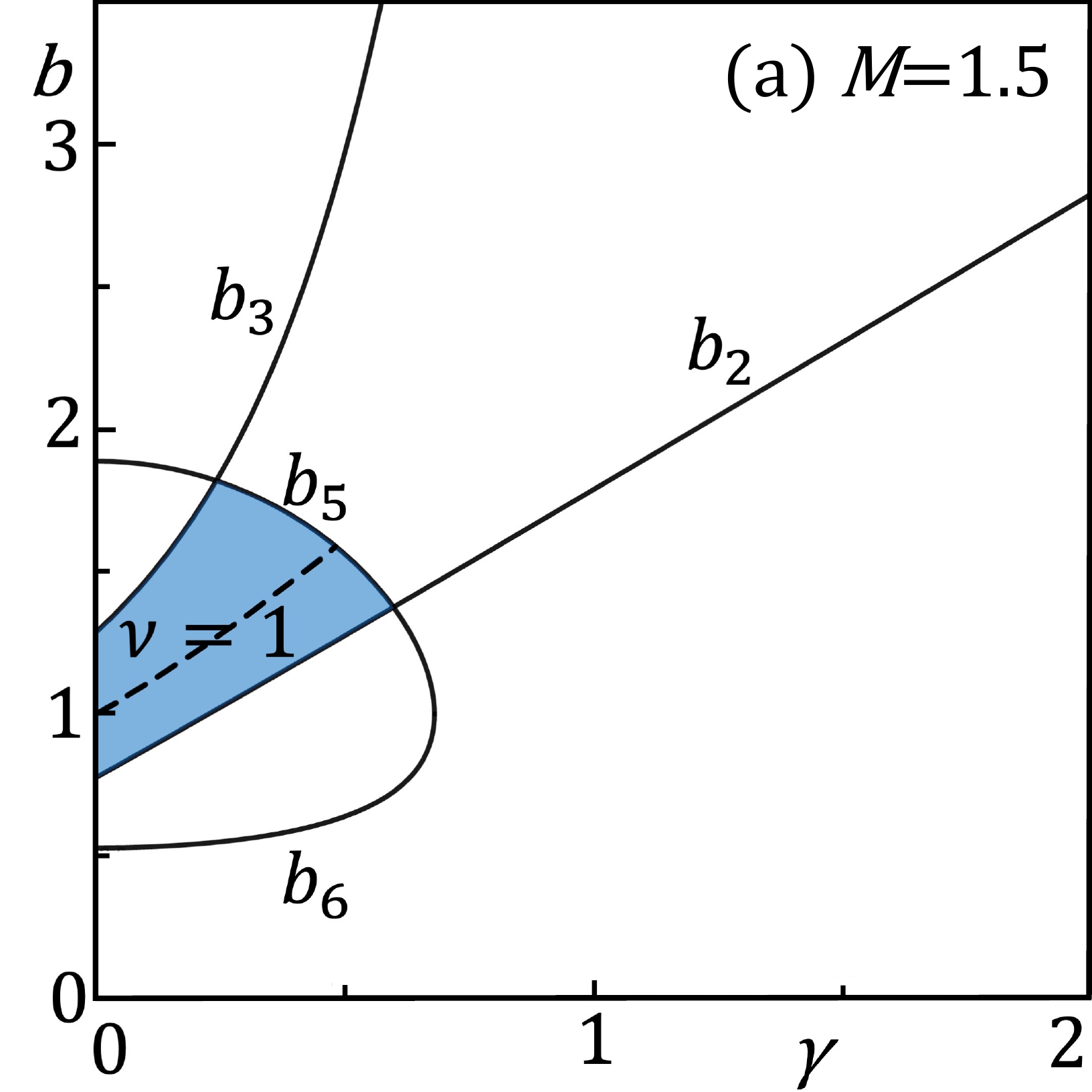}
\end{center}
\end{minipage}
\begin{minipage}{0.5\hsize}
\begin{center}
\hspace{-10mm}
\includegraphics[height=3.8cm]{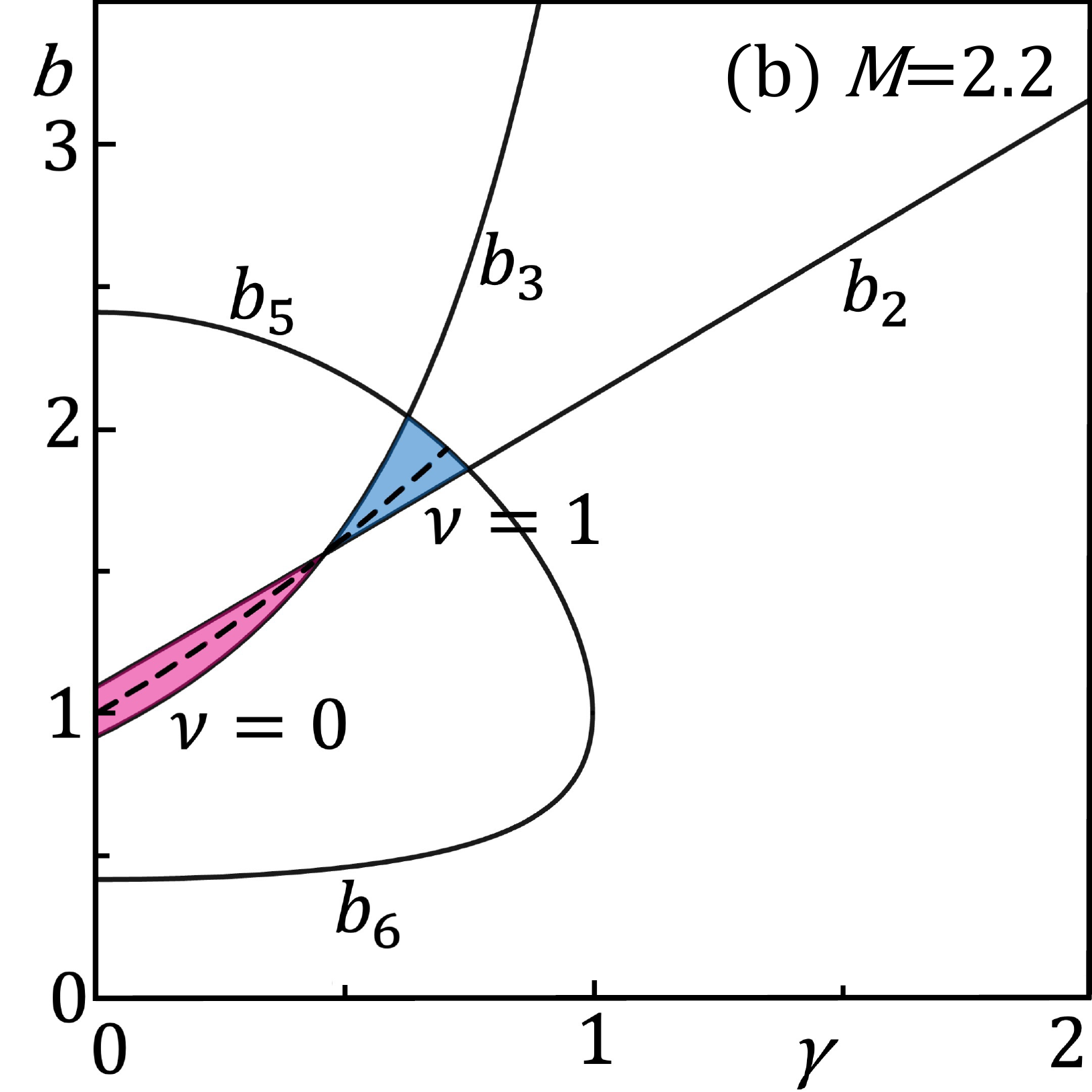}
\end{center}
\end{minipage}
\end{tabular}
\begin{tabular}{cc}
\begin{minipage}{0.5\hsize}
\begin{center}
\hspace{-10mm}
\includegraphics[height=3.8cm]{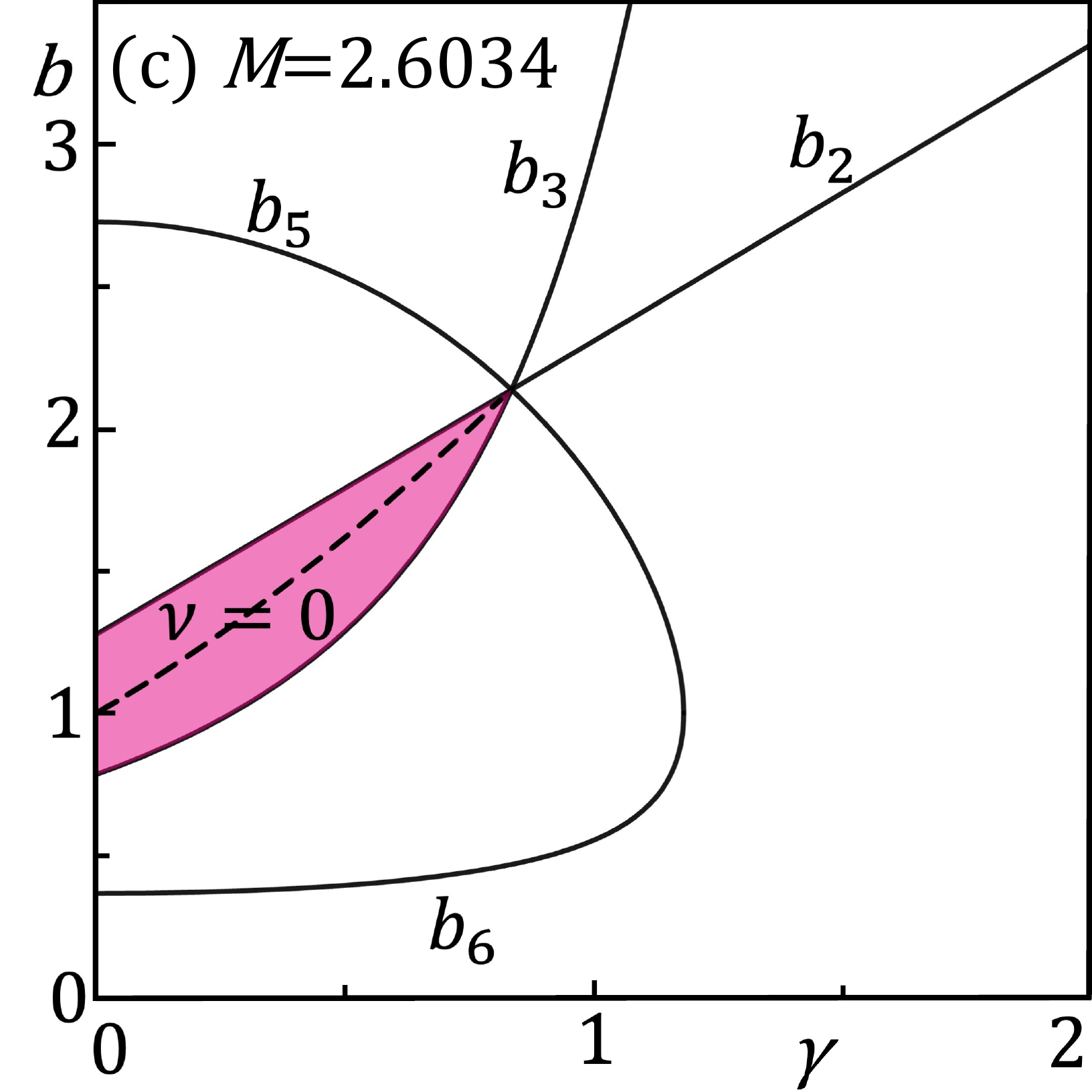}
\end{center}
\end{minipage}
\begin{minipage}{0.5\hsize}
\begin{center}
\hspace{-10mm}
\includegraphics[height=3.8cm]{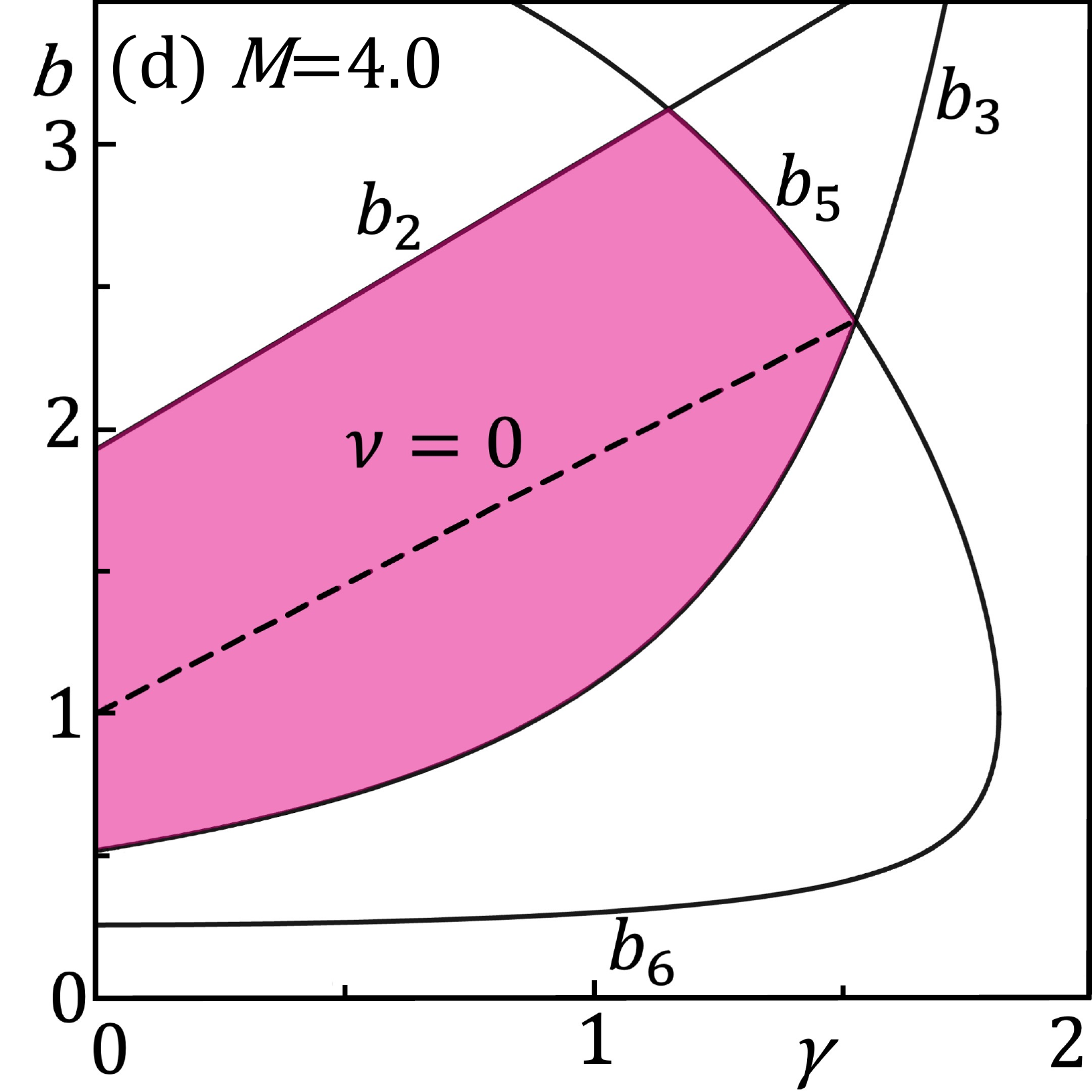}
\end{center}
\end{minipage}
\end{tabular}
\caption{
(Color online)
Distribution maps of $\nu$ in the $\gamma b$-plane for $\alpha = 1.2$
and $M =$ (a) $1.5$, (b) $2.2$, (c) $2.6034$, and (d) $4.0$.
Topologically trivial and nontrivial regions are respectively
designated as $\nu =0$ (mazenta) and $\nu =1$ (light blue),
outside of which is a gapless region.
In each figure, the dotted line represents a possible trajectory of
$b(\gamma)$.
}
\end{figure}

Figures~1 and 2 show $\nu$ calculated in the bulk geometry
as a function of $\gamma$ and $b$ for the given $\alpha$ and $M$.
Using these distribution maps of $\nu$,
let us consider which of the three phases appears in the boundary geometry.
In the scenario of non-Hermitian bulk--boundary
correspondence,~\cite{takane2} a phase realized in the boundary geometry
is governed by $\nu$ at $b(\gamma)$,
where $b(\gamma)$ is determined by the recipe given below.
In other words, the scenario predicts that $\nu(\gamma,b)$ with $b(\gamma)$ is
in one-to-one correspondence with a phase realized in the boundary geometry.
If $\nu = 1$ ($\nu = 0$), the nontrivial (trivial) phase
is realized in the boundary geometry.
We employ the recipe for determining $b(\gamma)$ given in Ref.~\citen{takane2}:
\begin{enumerate}
\item $b(0) = 1$ at the Hermitian limit of $\gamma = 0$.
\item With the exception described in (3), $b(\gamma)$ is allowed to cross
gap-closing lines only at a crossing point between the two.
\item If a peculiar gap closing is caused by the destabilization of
topological boundary states, $b(\gamma)$ in the nontrivial region
is determined in accordance with the peculiar gap closing.
\end{enumerate}
The first requirement is consistent with the original scenario
of the Hermitian bulk--boundary correspondence,~\cite{hatsugai,ryu2}
in which the pbc is imposed on the bulk geometry.

An explanation of the second requirement has been given
in Ref.~\citen{takane1}.
Here, we give it for self-containedness.
If $b(\gamma)$ crosses a gap-closing line, a zero-energy solution appears
at the crossing point, giving rise to a gapless spectrum in the bulk geometry.
Hence, to verify the bulk--boundary correspondence,
the spectrum in the boundary geometry must also be gapless at this point.
A single solution is insufficient to construct a general solution
compatible with the obc.~\cite{yao1,yokomizo1}
A crossing point between two $b_{i}$ yields two zero-energy solutions,
which should enable us to construct a zero-energy solution
in the boundary geometry under the obc.
We thus expect that $b(\gamma)$ is allowed to cross $b_{i}$
only at such a crossing point.
Although the second requirement does not uniquely determine
$b(\gamma)$ except at a crossing point,
$\nu(\gamma,b)$ is uniquely determined.~\cite{takane1,takane2}

A peculiar gap closing has been pointed out in Ref.~\citen{takane2}
for a non-Hermitian Chern insulator.
Here, we rephrase its explanation~\cite{takane2}
in the context of a non-Hermitian quantum spin-Hall insulator.
In the nontrivial phase in the boundary geometry,
the conduction and valence bands are linked by helical edge states.
If the helical edge states are destabilized by non-Hermiticity
and are thus transformed into bulk states,
the two bands are combined into one band
by a bridge of destabilized helical edge states.
This is referred to as a peculiar gap closing, which is intrinsic to
two- and three-dimensional non-Hermitian topological systems.
The third requirement is added to take this into account.~\cite{takane2}

Before giving an example of the peculiar gap closing,
we summarize the two important characteristics of helical edge states
in the boundary geometry of our model (see Appendix for details).
The first one is that the energy of the helical edge states must be real
as in the Hermitian limit of $\gamma = 0$ and therefore their spectrum
cannot deviate from the real axis although $\gamma \neq 0$.
Conversely, we can say that a helical edge state is destabilized
if its energy becomes complex.
The second one is that the wavefunction amplitude of a helical edge state
characterized by a wavenumber $k$
varies in the $x$- and $y$-directions with the same rate of increase:
\begin{align}
     \label{eq:b-CEs_bg}
  b = \sqrt{1+\left(\frac{\gamma}{\cos (ka)}\right)^{2}}
      + \frac{\gamma}{\cos (ka)} .
\end{align}
These two are characteristic features of our model and do not necessarily hold
in generic non-Hermitian quantum spin-Hall insulators.

Typical spectra in the boundary geometry of $100 \times 100$ sites
are shown in Fig.~3 to give an example of the peculiar gap closing.
Figure~3 shows spectra in the case of $\alpha = 0.2$ and $M = 2.4$
with (a) $\gamma = 1.01$ in the nontrivial phase
and (b) $\gamma = 1.07$ in the gapless phase.
In the nontrivial phase,
the conduction and valence bands are separated by a gap
and they are linked by helical edge states on the real axis.
In the gapless phase, the two bands are combined into one band by a bridge of
destabilized helical edge states off the real axis.
Note that the destabilized helical edge states are bulk states.

When the peculiar gap closing occurs in the boundary geometry,
helical edge states at zero energy are transformed into bulk states
at a transition point to the gapless phase.
Such helical edge states are characterized by
Eq.~(\ref{eq:b-CEs_bg}) at $k = 0$:
\begin{align}
    \label{eq:b-gamma}
  b = \sqrt{1+\gamma^{2}} + \gamma .
\end{align}
In accordance with the third requirement, we determine $b(\gamma)$
using Eq.~(\ref{eq:b-gamma}) in the nontrivial region.
At a crossing point of $b(\gamma)$ and a gap-closing line,
the bulk geometry becomes gapless.
The bulk--boundary correspondence requires that such a crossing must coincide
with the destabilization of helical edge states in the boundary geometry.
This is numerically verified in Sect.~5.
\begin{figure}[btp]
\begin{tabular}{cc}
\begin{minipage}{0.5\hsize}
\begin{center}
\hspace{-10mm}
\includegraphics[height=2.8cm]{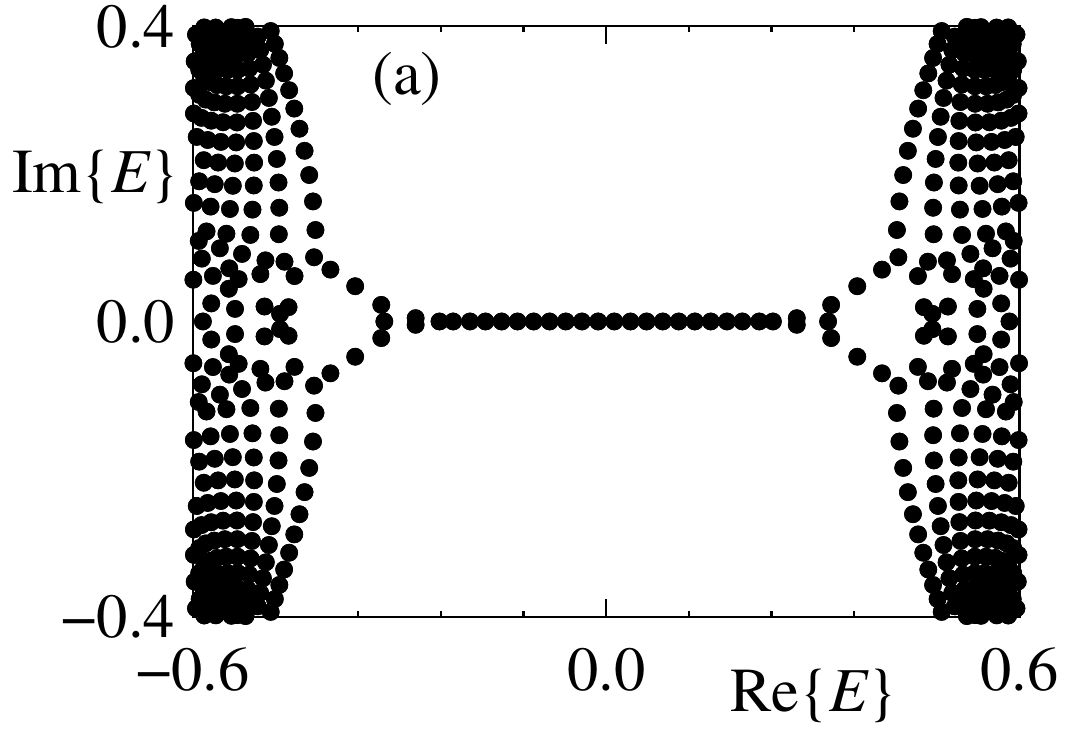}
\end{center}
\end{minipage}
\begin{minipage}{0.5\hsize}
\begin{center}
\hspace{-5mm}
\includegraphics[height=2.8cm]{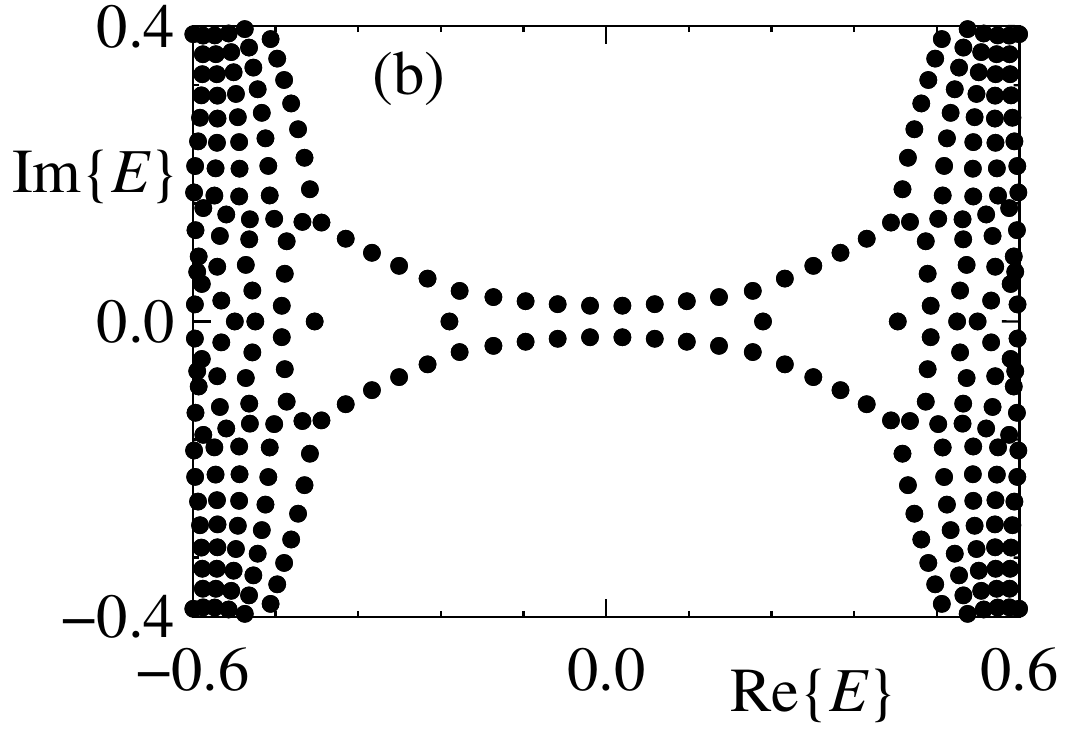}
\end{center}
\end{minipage}
\end{tabular}
\caption{
Spectra in the boundary geometry of $100 \times 100$ sites
in the case of $\alpha = 0.2$ and $M = 2.4$
with $\gamma =$ (a) $1.01$ and (b) $1.07$.
(a) shows the spectrum in the nontrivial phase
with helical edge states on the real axis.
(b) shows the spectrum in the gapless phase, where the conduction
and valence bands are combined into one band by a bridge of
the destabilized helical edge states off the real axis.
}
\end{figure}

Let us examine the bulk--boundary correspondence by considering
possible trajectories of $b(\gamma)$ that satisfy the three requirements.
Figures~1 and 2 show such $b(\gamma)$ in the cases of
$\alpha = 0.2$ and $\alpha = 1.2$, respectively.
The trajectories in the nontrivial regions ($\nu = 1$)
are determined by the third requirement,
whereas the trajectories in the trivial regions ($\nu = 0$)
are determined by the first and second requirements.
A phase realized in the boundary geometry is governed by $\nu$ at $b(\gamma)$.
For example, Fig.~1(c) predicts that the phase realized
in the boundary geometry in the case of $\alpha = 0.2$ and $M = 2.4$
starts from the trivial phase ($\nu = 0$) at $\gamma = 0$,
changes to the nontrivial phase ($\nu = 1$) with increasing $\gamma$,
and finally enters the gapless phase.

We determine three phase boundaries from Fig.~1.
The first one is between the trivial region ($\nu = 0$)
and the nontrivial region ($\nu = 1$).
This is at $M = 2$ in the Hermitian limit of $\gamma = 0$.
In the non-Hermitian regime of $\gamma > 0$,
this is determined by the condition $b_{2}=b_{3}$,
as can be seen in Figs.~1(b) and 1(c).
Solving $b_{2}=b_{3}$, we find
\begin{align}
     \label{eq:PB_tri-nontri}
  M = 2\sqrt{1 + \gamma^{2}}
\end{align}
when $2 < M < 2\sqrt{2}(1-\alpha^{2})^{-\frac{1}{2}}$.
The second one is between the trivial region ($\nu = 0$)
and the gapless region.
As can be seen in Figs.~1(d), this is determined by the condition
$b_{1} = b_{3}$, resulting in
\begin{align}
     \label{eq:PB_tri-gapless}
  M = \sqrt{2(1+\alpha^{2})}\gamma + \sqrt{2(1-\alpha^{2})}
\end{align}
when $2\sqrt{2}(1-\alpha^{2})^{-\frac{1}{2}} < M$.
The third one is between the nontrivial region ($\nu = 1$)
and the gapless region.
In the range of $M < 2(1+\alpha^{2})(1-\alpha^{2})^{-\frac{1}{2}}$,
this is determined by the condition $b(\gamma) = b_{5}$,
as can be seen in Fig.~1(a), resulting in
\begin{align}
    \label{eq:nontrivial-gapless1}
  M = 2\sqrt{1+\alpha^{2}}\gamma .
\end{align}
The phase boundary in the range of
$2(1+\alpha^{2})(1-\alpha^{2})^{-\frac{1}{2}} < M
< 2\sqrt{2}(1-\alpha^{2})^{-\frac{1}{2}}$
is determined by the condition $b(\gamma) = b_{0}$,
as can be seen in Fig.~1(c), resulting in
\begin{align}
    \label{eq:nontrivial-gapless2}
  \gamma = \sqrt{\frac{1+\alpha^{2}}{1-\alpha^{2}}} .
\end{align}
These results are justified when $0 \le \alpha < \sqrt{\sqrt{2}-1}$.
The phase diagram in the boundary geometry
in the case of $\alpha = 0.2$ is shown in Fig.~4(a).
\begin{figure}[btp]
\begin{tabular}{cc}
\begin{minipage}{0.5\hsize}
\begin{center}
\hspace{-10mm}
\includegraphics[height=3.5cm]{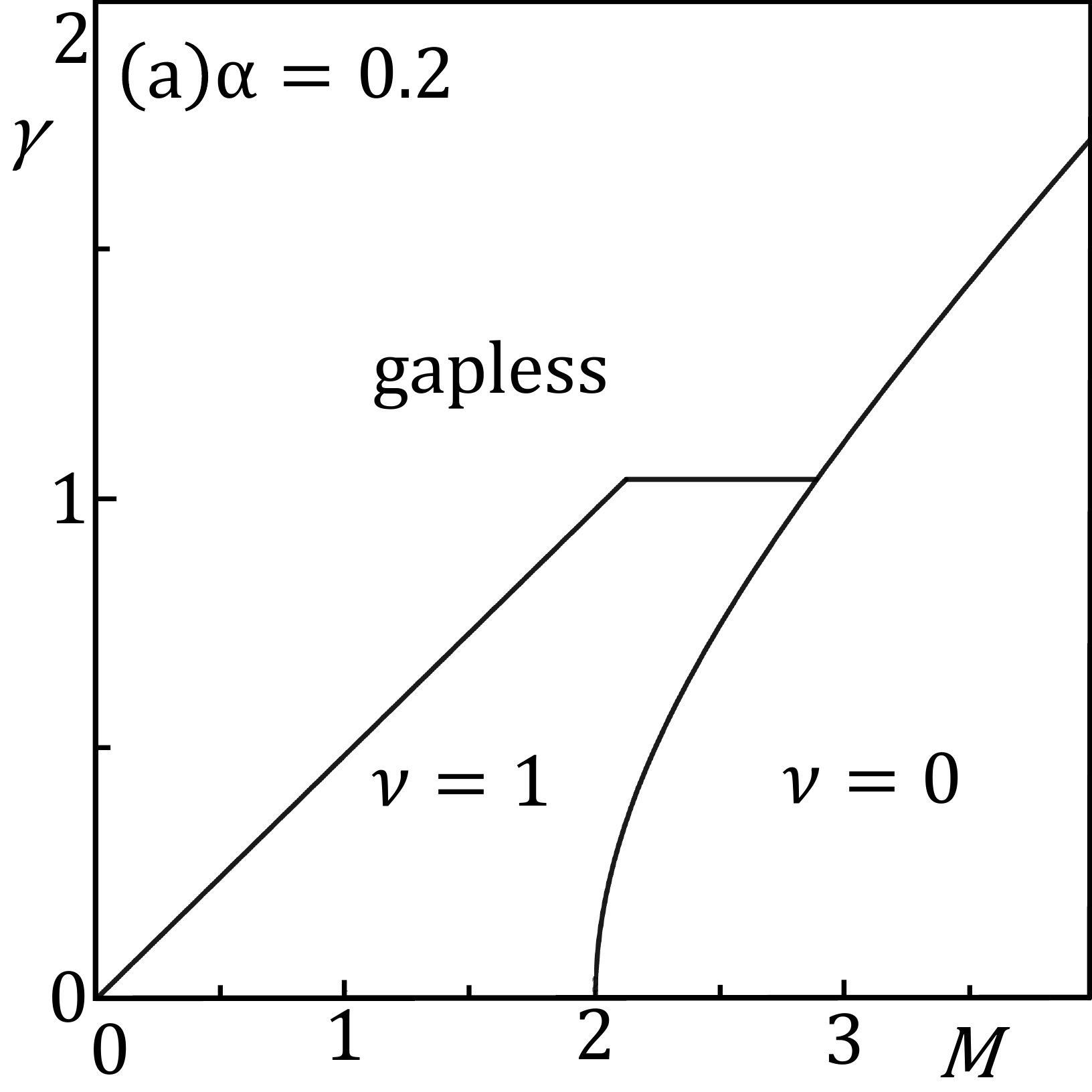}
\end{center}
\end{minipage}
\begin{minipage}{0.5\hsize}
\begin{center}
\hspace{-10mm}
\includegraphics[height=3.5cm]{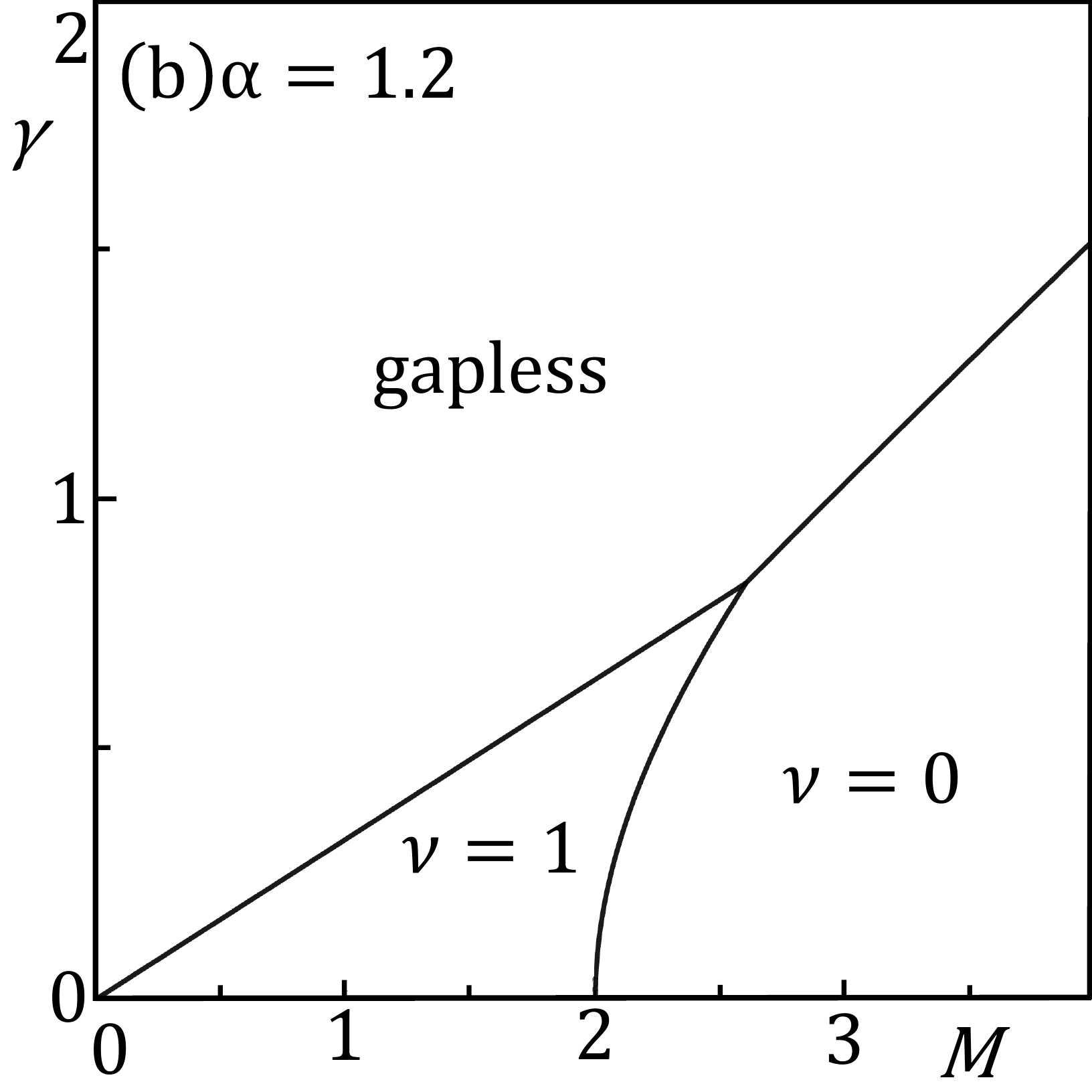}
\end{center}
\end{minipage}
\end{tabular}
\caption{
Phase diagrams in the boundary geometry with $\alpha =$
(a) $0.2$ and (b) $1.2$, where the regions designated by $\nu = 0$
and $\nu = 1$ correspond to the topologically trivial and nontrivial regions,
respectively, outside of which is the gapless region.
}
\end{figure}

We next determine the three phase boundaries from Fig.~2.
The phase boundary between the trivial region ($\nu = 0$)
and the nontrivial region ($\nu = 1$) is again determined by the condition
$b_{2}=b_{3}$, which yields the same result, Eq.~(\ref{eq:PB_tri-nontri}),
as in the case of $0 \le \alpha < \sqrt{\sqrt{2}-1}$.
Equation~(\ref{eq:PB_tri-nontri}) holds in the range of
$2 \le M < 2\sqrt{1+\alpha^{-2}}$ in this case.
The phase boundary between the trivial region ($\nu = 0$)
and the gapless region is determined by the condition
$b_{3} = b_{5}$ as is seen in Fig.~2(d).
The resulting boundary,
which appears in the range of $2\sqrt{1+\alpha^{-2}} < M$,
is difficult to give in an analytic form.
The phase boundary between the nontrivial region ($\nu = 1$)
and the gapless region is determined by the condition
$b(\gamma) = b_{5}$ as can be seen in Figs.~2(a) and 2(b).
Solving $b(\gamma)=b_{5}$,
we find the same result, Eq.~(\ref{eq:nontrivial-gapless1}),
as in the case of $0 \le \alpha < \sqrt{\sqrt{2}-1}$.
Equation~(\ref{eq:nontrivial-gapless1}) holds
in the range of $M < 2\sqrt{1+\alpha^{-2}}$ in this case.
These results are justified when $\sqrt{\sqrt{2}-1} \le \alpha$.
The phase diagram in the boundary geometry
in the case of $\alpha = 1.2$ is shown in Fig.~4(b).

\section{Numerical Results}

We examine the phase boundaries shown in Fig.~4
by comparing them with spectra in the boundary geometry under the obc.
If a right eigenvector is written in the form of
Eq.~(\ref{eq:right-eigenvec}), the obc is expressed as
\begin{align}
 & |\psi^{R}(m,0)\rangle = |\psi^{R}(m,N+1)\rangle = |\psi^{R}(0,n)\rangle
      \nonumber \\
 & \hspace{5mm}
   = |\psi^{R}(N+1,n)\rangle = \bold{0}
\end{align}
for $0 \le m, n \le N+1$.
To determine the spectrum of $H$ for a given set of parameters,
we introduce $\tilde{H} = \Lambda^{-1}H\Lambda$ with
\begin{align}
   \label{eq:def-b_similarity}
   \Lambda
   = \sum_{m,n} |m,n \rangle b_{\rm S}^{m+n} \langle m,n| ,
\end{align}
where $\Lambda$ represents a similarity transformation.
Because the spectrum of $H$ is invariant under a similarity transformation,
we can determine the spectrum of $H$ by computing eigenvalues of $\tilde{H}$.
The accuracy of computation is improved
if we appropriately set the value of $b_{\rm S}$.
In determining spectra near a phase boundary, $b_{\rm S}$ in $\Lambda$
should be set equal to $b(\gamma)$ at a crossing point
that corresponds to the phase boundary (see Figs.~1 and 2).
An actual value of $b_{\rm S}$ used to determine each spectrum is given
in the corresponding captions of Figs.~5 and 6.
We use numerical libraries of z-Pares~\cite{sakurai,futamura}
with MUMPS~\cite{amestoy} to compute the eigenvalues.

\begin{figure}[btp]
\begin{tabular}{cc}
\begin{minipage}{0.5\hsize}
\begin{center}
\hspace{-10mm}
\includegraphics[height=3.0cm]{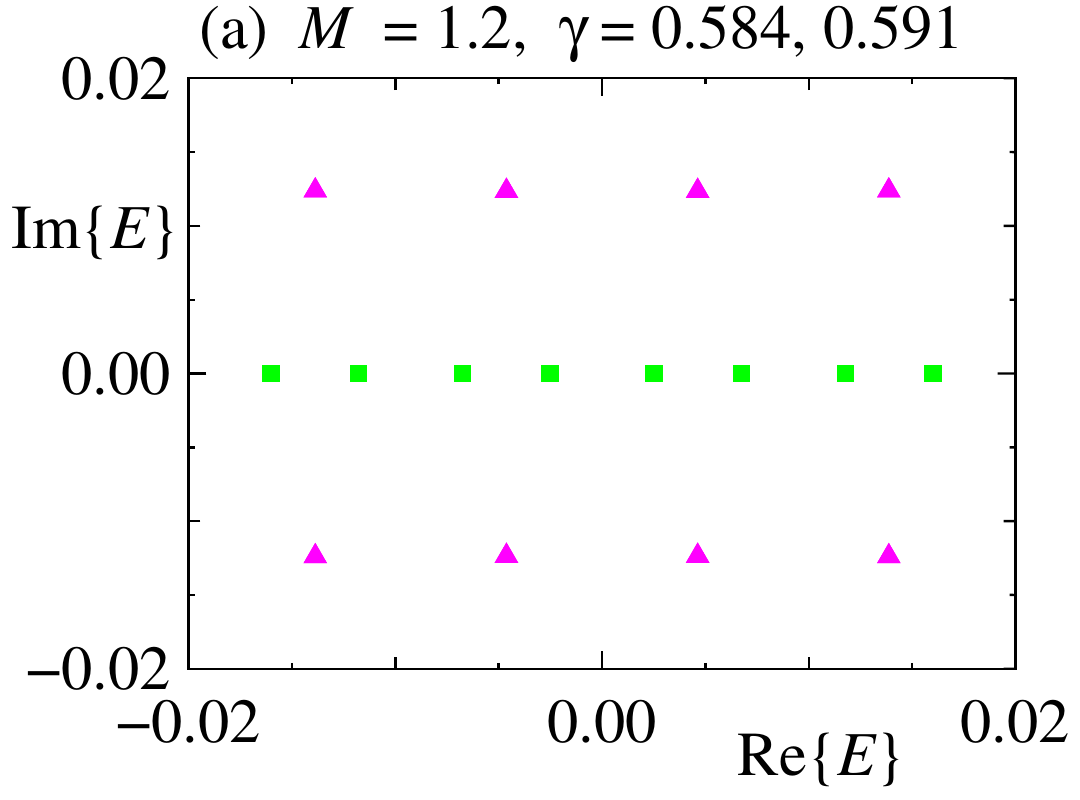}
\end{center}
\end{minipage}
\begin{minipage}{0.5\hsize}
\begin{center}
\hspace{-5mm}
\includegraphics[height=3.0cm]{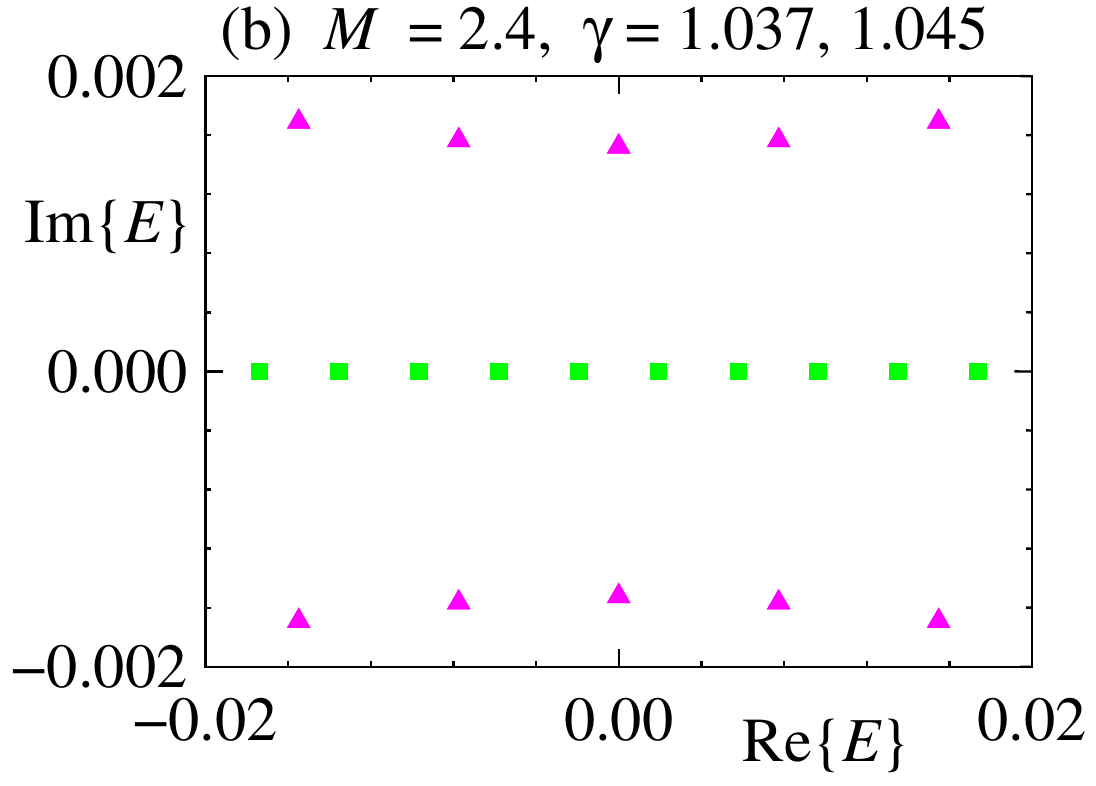}
\end{center}
\end{minipage}
\end{tabular}
\begin{tabular}{cc}
\begin{minipage}{0.5\hsize}
\begin{center}
\hspace{-10mm}
\includegraphics[height=3.0cm]{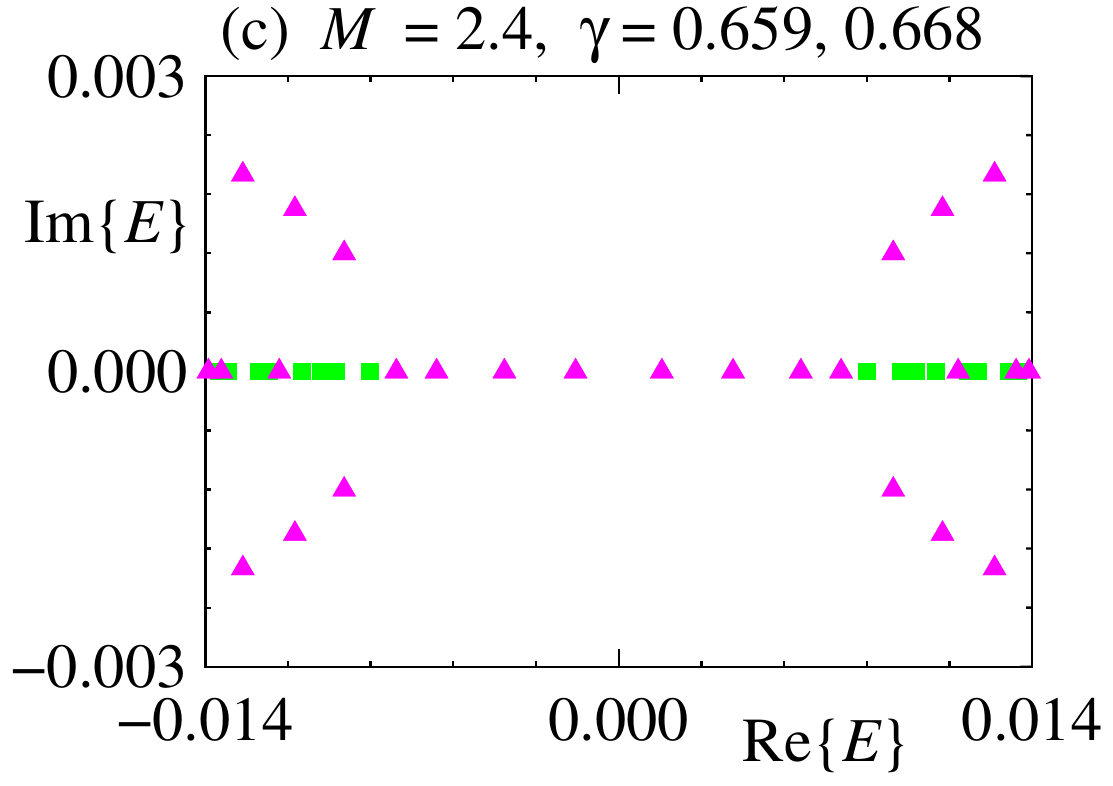}
\end{center}
\end{minipage}
\begin{minipage}{0.5\hsize}
\begin{center}
\hspace{-5mm}
\includegraphics[height=3.0cm]{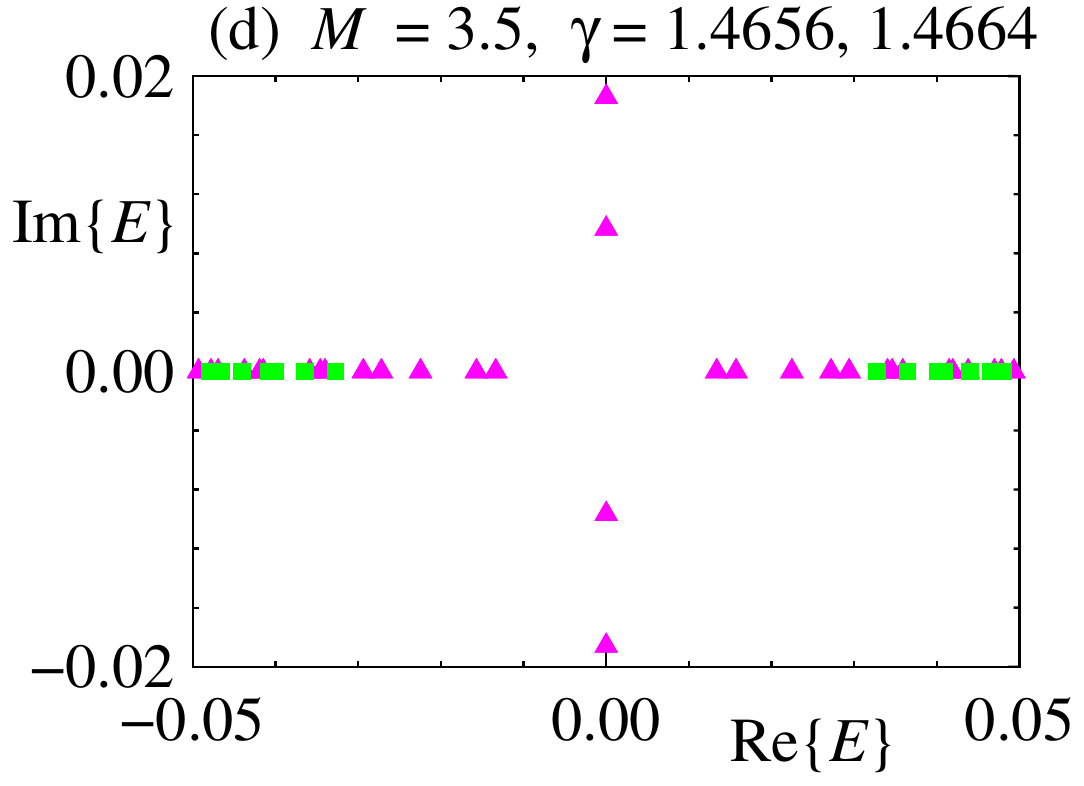}
\end{center}
\end{minipage}
\end{tabular}
\caption{
(Color online)
Spectra in the boundary geometry of $N \times N$ sites at $\alpha = 0.2$.
The parameters are
(a) $M = 1.2$, $N = 400$, and $b_{\rm S} = 1.74858$
with $\gamma = 0.584$ (squares) and $0.591$ (triangles),
where the phase boundary is at $\gamma_{\rm c} \approx 0.588$,
(b) $M = 2.4$, $N = 600$, and $b_{\rm S} = 2.48400$
with $\gamma = 1.037$ (squares) and $1.045$ (triangles),
where the phase boundary is at $\gamma_{\rm c} \approx 1.041$,
(c) $M = 2.4$, $N = 1000$, and $b_{\rm S} = 1.86330$
with $\gamma = 0.659$ (squares) and $0.668$ (triangles),
where the phase boundary is at $\gamma_{\rm c} = 0.663$,
and (d) $M = 3.5$, $N = 300$, and $b_{\rm S} = 2.48400$
with $\gamma = 1.4656$ (squares) and $1.4664$ (triangles),
where the phase boundary is at $\gamma_{\rm c} \approx 1.466$.}
\end{figure}
Let us first examine the phase diagram
in the case of $\alpha = 0.2$ [i.e., Fig.~4(a)].
Figure~5 shows typical spectra in this case, where only eigenvalues
near the band center [i.e., $E = (0,0)$] are shown.
The phase boundary between the nontrivial and gapless regions is examined
in Fig.~5(a) with $M = 1.2$ and Fig.~5(b) with $M = 2.4$.
These figures focus on a small central region in the gap
between the conduction and valence bands.
In the case of $M = 1.2$,
the phase boundary is expected at $\gamma_{\rm c} \approx 0.588$.
The spectrum at $\gamma = 0.584$ (squares) consists of eigenvalues
that are almost equally distributed on the real axis.
They represent the helical edge states.
In contrast, in the spectrum at $\gamma = 0.591$ (triangles),
eigenvalues are split into two branches off the real axis, showing that
the helical edge states are destabilized and transformed into bulk states.
That is, a peculiar gap closing occurs:
the conduction and valence bands are combined into one band
by a bridge of destabilized helical edge states.
These results are consistent with $\gamma_{\rm c} \approx 0.588$,
at which the system changes from the nontrivial phase to the gapless phase.
In the case of $M = 2.4$,
the phase boundary is expected at $\gamma_{\rm c} \approx 1.041$.
The spectrum at $\gamma = 1.037$ (squares) consists of eigenvalues
that are almost equally distributed on the real axis,
whereas the spectrum at $\gamma = 1.045$ (triangles) consists of eigenvalues
that are split into two branches off the real axis.
These results are consistent with $\gamma_{\rm c} \approx 1.041$,
at which the system changes from the nontrivial phase to the gapless phase.
The phase boundary between the trivial and nontrivial regions is examined
in Fig.~5(c) with $M = 2.4$.
The phase boundary is expected at $\gamma_{\rm c} \approx 0.663$.
The spectrum at $\gamma = 0.659$ (squares) has a small gap
between the conduction and valence bands without helical edge states.
In contrast, the gap in the spectrum at $\gamma = 0.668$ (triangles) is
filled with eigenvalues that are almost equally distributed on the real axis.
They represent the helical edge states.
These results are consistent with $\gamma_{\rm c} \approx 0.663$,
at which the system changes from the trivial phase to the nontrivial phase.
The phase boundary between the trivial and gapless regions is examined
in Fig.~5(d) with $M = 3.5$.
The phase boundary is expected at $\gamma_{\rm c} \approx 1.466$.
The spectrum at $\gamma = 1.4656$ (squares) has a small gap
between the conduction and valence bands without helical edge states.
In contrast, the spectrum at $\gamma = 1.4664$ (triangles) is gapless.
These results are consistent with $\gamma_{\rm c} \approx 1.466$,
at which the system changes from the trivial phase to the gapless phase.

\begin{figure}[btp]
\begin{tabular}{cc}
\begin{minipage}{0.5\hsize}
\begin{center}
\hspace{-10mm}
\includegraphics[height=3.0cm]{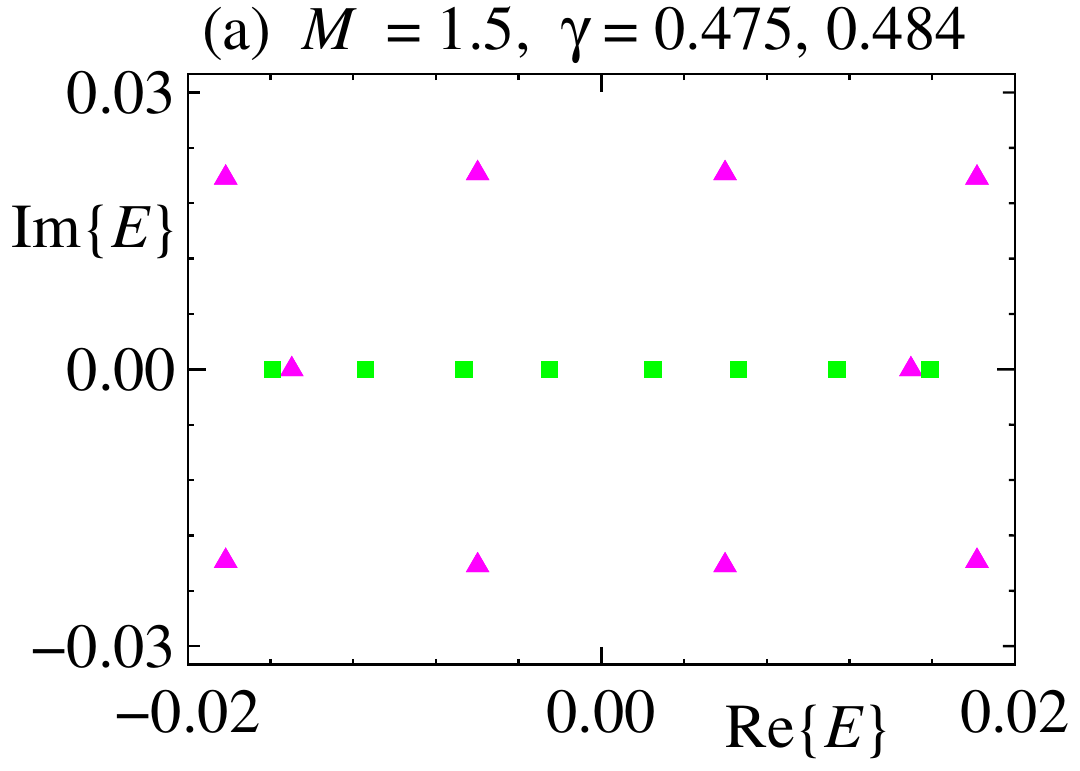}
\end{center}
\end{minipage}
\begin{minipage}{0.5\hsize}
\begin{center}
\hspace{-5mm}
\includegraphics[height=3.0cm]{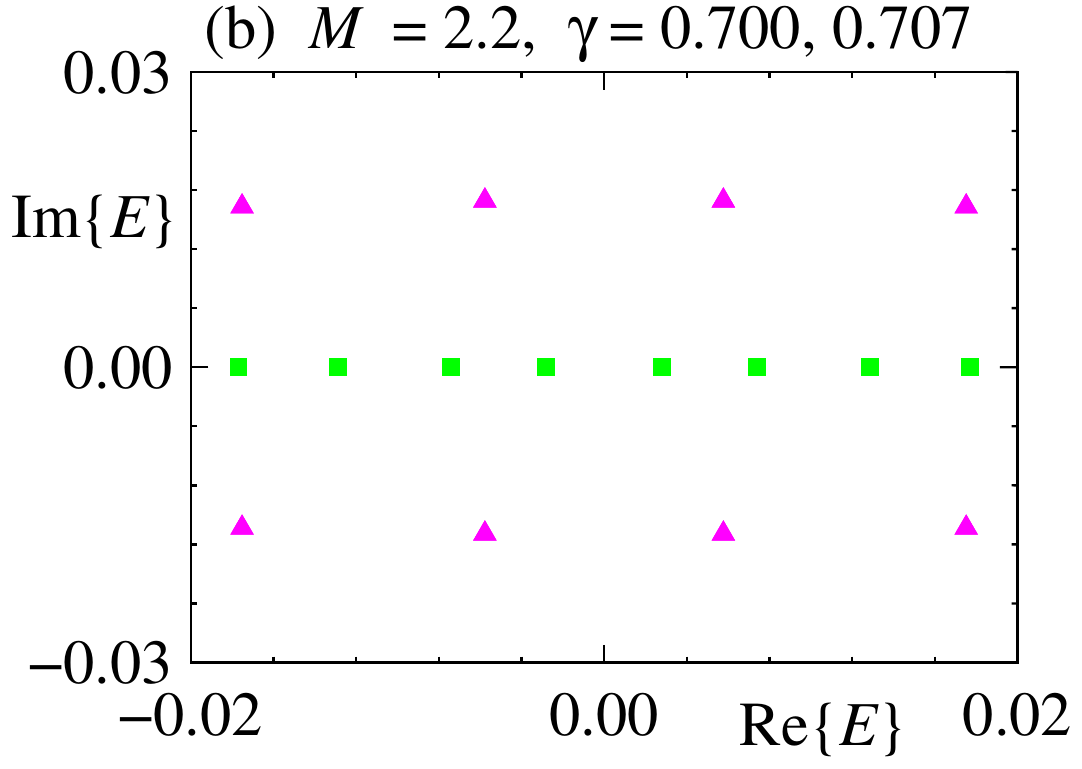}
\end{center}
\end{minipage}
\end{tabular}
\begin{tabular}{cc}
\begin{minipage}{0.5\hsize}
\begin{center}
\hspace{-10mm}
\includegraphics[height=3.0cm]{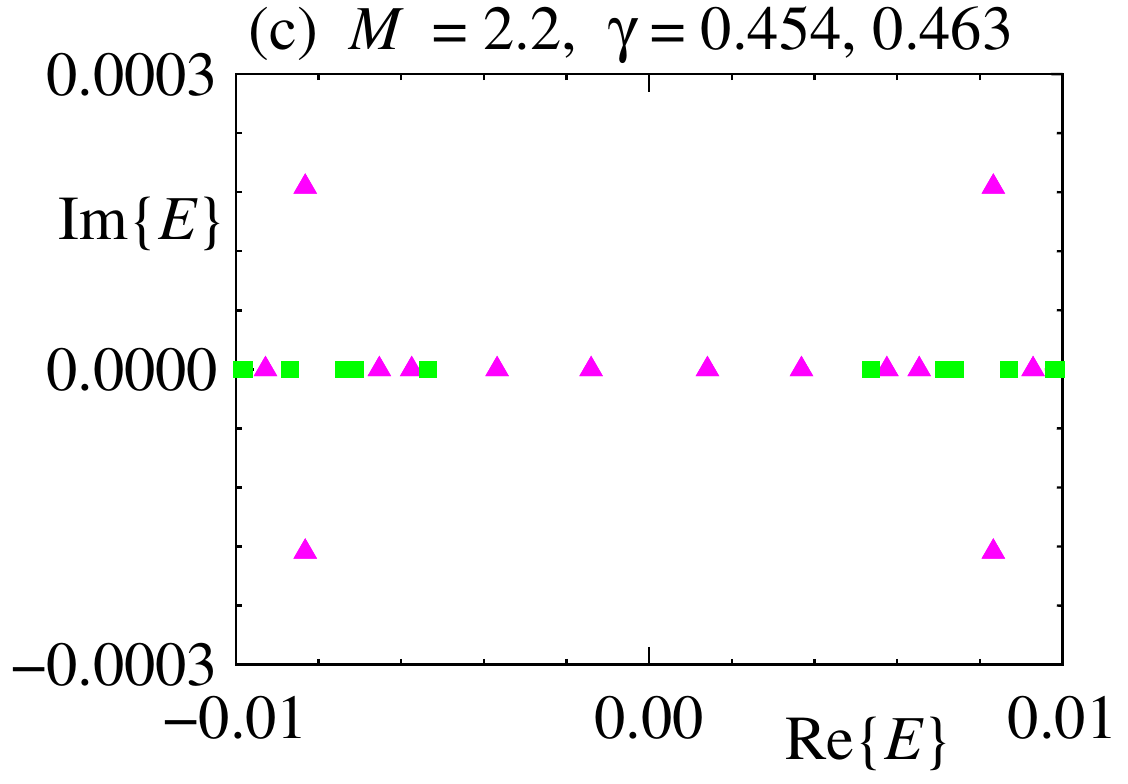}
\end{center}
\end{minipage}
\begin{minipage}{0.5\hsize}
\begin{center}
\hspace{-5mm}
\includegraphics[height=3.0cm]{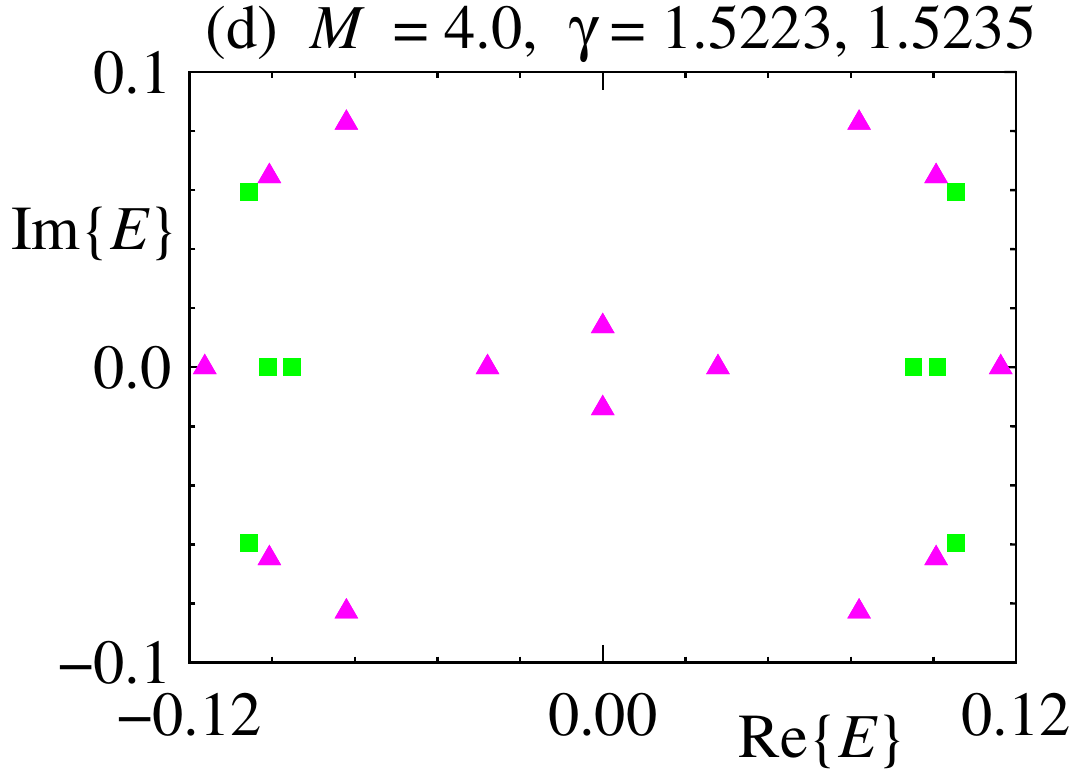}
\end{center}
\end{minipage}
\end{tabular}
\caption{
(Color online)
Spectra in the boundary geometry of $N \times N$ sites at $\alpha = 1.2$.
The parameters are
(a) $M = 1.5$, $N = 600$, and $b_{\rm S} = 1.58943$
with $\gamma = 0.475$ (squares) and $0.484$ (triangles),
where the phase boundary is at $\gamma_{\rm c} \approx 0.480$,
(b) $M = 2.2$, $N = 600$, and $b_{\rm S} = 1.92727$
with $\gamma = 0.700$ (squares) and $0.707$ (triangles),
where the phase boundary is at $\gamma_{\rm c} \approx 0.704$,
(c) $M = 2.2$, $N = 1600$, and $b_{\rm S} = 1.55825$
with $\gamma = 0.454$ (squares) and $0.463$ (triangles),
where the phase boundary is at $\gamma_{\rm c} = 0.458$,
and (d) $M = 4.0$, $N = 400$, and $b_{\rm S} = 2.37900$
with $\gamma = 1.5223$ (squares) and $1.5235$ (triangles),
where the phase boundary is at $\gamma_{\rm c} \approx 1.5229$.
}
\end{figure}
Let us next examine the phase diagram
in the case of $\alpha = 1.2$ [i.e., Fig.~4(b)].
Figure~6 shows typical spectra in this case,
where only eigenvalues near the band center [i.e., $E = (0,0)$] are shown.
The phase boundary between the nontrivial and gapless regions is examined
in Fig.~6(a) with $M = 1.5$ and Fig.~6(b) with $M = 2.2$.
These figures focus on a small central region in the gap
between the conduction and valence bands.
In the case of $M = 1.5$,
the phase boundary is expected at $\gamma_{\rm c} \approx 0.480$.
The spectrum at $\gamma = 0.475$ (squares) consists of eigenvalues
that are almost equally distributed on the real axis.
They represent the helical edge states.
In contrast, in the spectrum at $\gamma = 0.484$ (triangles),
most of the eigenvalues are off the real axis, showing that
the helical edge states are destabilized and transformed into bulk states.
That is, a peculiar gap closing occurs.
These results are consistent with $\gamma_{\rm c} \approx 0.480$,
at which the system changes from the nontrivial phase to the gapless phase.
In the case of $M = 2.2$,
the phase boundary is expected at $\gamma_{\rm c} \approx 0.704$.
The spectrum at $\gamma = 0.700$ (squares) consists of eigenvalues
that are almost equally distributed on the real axis,
whereas most of the eigenvalues
in the spectrum at $\gamma = 0.707$ (triangles) are off the real axis.
These results are consistent with $\gamma_{\rm c} \approx 0.704$,
at which the system changes from the nontrivial phase to the gapless phase.
The phase boundary between the trivial and nontrivial regions is examined
in Fig.~6(c) with $M = 2.2$.
The phase boundary is expected at $\gamma_{\rm c} \approx 0.458$.
The spectrum at $\gamma = 0.454$ (squares) has a small gap
between the conduction and valence bands without helical edge states.
In contrast, the gap in the spectrum at $\gamma = 0.463$ (triangles) is
filled with eigenvalues that are almost equally distributed on the real axis.
They represent the helical edge states.
These results are consistent with $\gamma_{\rm c} \approx 0.458$,
at which the system changes from the trivial phase to the nontrivial phase.
The phase boundary between the trivial and gapless regions is examined
in Fig.~6(d) with $M = 4.0$.
The phase boundary is expected at $\gamma_{\rm c} \approx 1.5229$.
The spectrum at $\gamma = 1.5223$ (squares) has a small gap
between the conduction and valence bands without helical edge states.
In contrast, the spectrum at $\gamma = 1.5235$ (triangles) is gapless.
These results are consistent with $\gamma_{\rm c} \approx 1.5229$,
at which the system changes from the trivial phase to the gapless phase.

\section{Summary and Discussion}

The original scenario of non-Hermitian bulk--boundary
correspondence~\cite{imura1,imura2}
is targeted to one-dimensional non-Hermitian topological systems.
A revised scenario~\cite{takane2} has been proposed to describe
the non-Hermitian bulk--boundary correspondence
in a two-dimensional Chern insulator.
In this study, we examined its applicability to another prototypical
two-dimensional topological system of
a non-Hermitian quantum spin-Hall insulator.
We show that the revised scenario properly describes
the bulk--boundary correspondence in the non-Hermitian quantum spin-Hall
insulator including the destabilization of helical edge states.
Indeed, a phase diagram derived from the bulk--boundary correspondence
is consistent with spectra in the boundary geometry.
This suggests that the revised scenario of non-Hermitian
bulk--boundary correspondence is applicable to a wide variety of
non-Hermitian topological systems.

\section*{Acknowledgment}

This work was supported by JSPS KAKENHI Grant Number JP21K03405.

\section*{Appendix}

\renewcommand{\thefigure}{\Alph{section}-\arabic{table}}
\setcounter{section}{1}
\setcounter{table}{1}

We elucidate two important characteristics of helical edge states
by considering the geometries shown in Fig.~A-1.
To do this, we apply the argument given in Ref.~\citen{takane2}.
\begin{figure}[btp]
\begin{center}
\includegraphics[height=2.6cm]{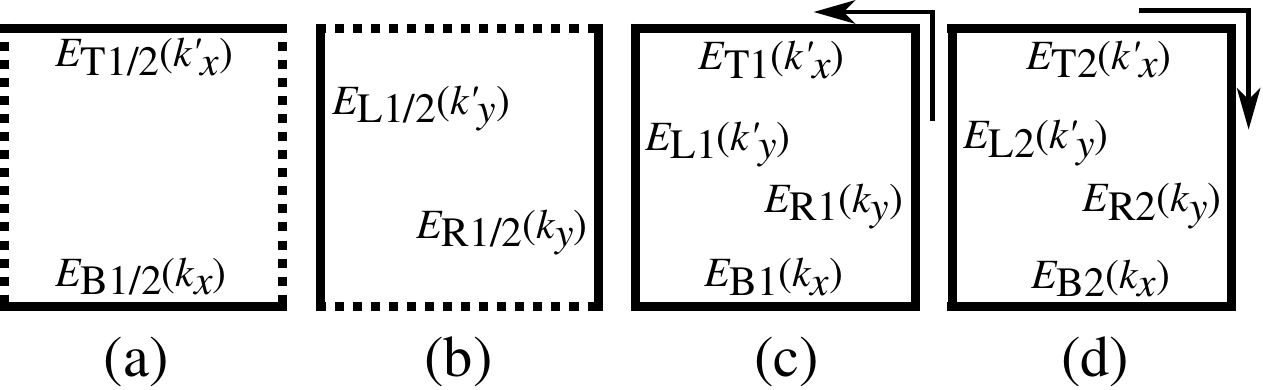}
\end{center}
\caption{
Square geometries in (a) and (b) are used to consider helical edge states.
The obc is imposed on edges denoted by solid lines
and the mpbc is imposed on a pair of edges denoted by dotted lines.
(c) and (d) represent how $E_{{\rm B}q}$, $E_{{\rm T}q}$,
$E_{{\rm L}q}$, and $E_{{\rm R}q}$ ($q = 1$, $2$)
are combined to give helical edge states
propagating in the anticlockwise and clockwise directions, respectively.
}
\end{figure}

We consider a right eigenvector written in the form of
Eq.~(\ref{eq:right-eigenvec}) with Eq.~(\ref{eq:psi^R}) in the system
[see Fig.~A-1(a)] under the obc in the $y$-direction,
\begin{align}
  \varphi^{R}(m,0) & = \varphi^{R}(m,N+1) = 0 ,
\end{align}
and the mpbc in the $x$-direction,
\begin{align}
  \varphi^{R}(m+N,n) & = b^{N} \varphi^{R}(m,n) .
\end{align}
To obtain fundamental solutions, we set
\begin{align}
  \varphi^{R}(m,n) = \beta_{x}^{m}\rho^{n} ,
\end{align}
where $\rho$ characterizes the penetration of helical edge states
in the $y$-direction and $\beta_{x} = be^{ik_{x}a}$.
We decompose the reduced Hamiltonian as
$h = h^{(0)} + h^{(1)}$ with
\begin{align}
   h^{(0)}
 & = \left[ \begin{array}{cc}
               \tilde{\eta}_{y}\tau_{y}+\tilde{\eta}_{z}\tau_{z} &
               -i \alpha\tilde{\eta}_{y}\tau_{x} \\
               i\alpha\tilde{\eta}_{y}\tau_{x} &
               \tilde{\eta}_{y}\tau_{y}+\tilde{\eta}_{z}\tau_{z}
             \end{array}
      \right] ,
           \\
   h^{(1)}
 & = \left[ \begin{array}{cc}
               \eta_{x}\tau_{x} & \alpha\eta_{x}\tau_{x} \\
               \alpha\eta_{x}\tau_{x} & -\eta_{x}\tau_{x}
             \end{array}
      \right] ,
\end{align}
where $\eta_{x}$ is given in Eq.~(\ref{eq:def-eta_x-mod}) and
\begin{align}
   \tilde{\eta}_{y}(\rho)
   & = \frac{1}{2i}\left(\rho-\rho^{-1}\right) + i\gamma,
     \\
   \tilde{\eta}_{z}(\rho)
   & = M - \frac{1}{2}\left(\beta_{x}+\beta_{x}^{-1}\right)
         - \frac{1}{2}\left(\rho+\rho^{-1}\right) .
\end{align}

Assuming that $N$ is sufficiently large, we construct right eigenvectors
of $h$ by superposing two solutions satisfying
\begin{align}
      \label{eq:eve-localized}
   h^{(0)}(\rho_{p})|\psi_{p}^{R}\rangle
 =  E_{\bot}|\psi_{p}^{R}\rangle
\end{align}
with $p = \pm$ as
\begin{align}
       \label{eq:phi-R-localized}
  |\Psi^{R}\rangle
   = \sum_{m}\beta_{x}^{m}
     \sum_{n=0}^{N+1}
     |m,n \rangle \cdot
     \left( c_{+}\rho_{+}^{n}|\psi_{+}^{R}\rangle
            + c_{-}\rho_{-}^{n}|\psi_{-}^{R}\rangle \right)   ,
\end{align}
where $c_{+}$ and $c_{-}$ are arbitrary constants.
The obc is satisfied
if $|\rho_{+}| < 1$ and $|\rho_{-}| < 1$ in addition to
\begin{align}
      \label{eq:condition1}
  c_{+}|\psi_{+}^{R}\rangle  + c_{-}|\psi_{-}^{R}\rangle
  = \bold{0} .
\end{align}
In this case, $|\Psi^{R}\rangle$ represents an eigenvector
localized near the bottom edge.
The obc is also satisfied
if $|\rho_{+}| > 1$ and $|\rho_{-}| > 1$ in addition to
\begin{align}
      \label{eq:condition2}
  c_{+} \rho_{+}^{N+1}|\psi_{+}^{R}\rangle
  + c_{-} \rho_{-}^{N+1}|\psi_{-}^{R}\rangle
  = \bold{0} .
\end{align}
In this case, $|\Psi^{R}\rangle$ represents an eigenvector
localized near the top edge.
The trial function given in Eq.~(\ref{eq:phi-R-localized}) can satisfy
Eq.~(\ref{eq:condition1}), or Eq.~(\ref{eq:condition2}), only if
\begin{align}
   \label{eq:v1=v2}
   |\psi_{+}^{R}\rangle = |\psi_{-}^{R}\rangle
\end{align}
with $\rho_{+} \neq \rho_{-}$,
which results in $E_{\bot} = 0$.~\cite{takane3}
To show this, we rewrite Eq.~(\ref{eq:eve-localized}) as
\begin{align}
     \label{eq:eve-appendix}
      \left[ \begin{array}{cc}
               \hspace{-1.5mm}
               \frac{\tilde{\eta}_{z}(\rho_{p})\tau_{z}-E_{\bot}\tau_{0}}
                    {-i\tilde{\eta}_{y}(\rho_{p})}+i\tau_{y}
               & \alpha\tau_{x} \\
               -\alpha\tau_{x} &
               \hspace{-1.0mm}
               \frac{\tilde{\eta}_{z}(\rho_{p})\tau_{z}-E_{\bot}\tau_{0}}
                    {-i\tilde{\eta}_{y}(\rho_{p})}+i\tau_{y}
               \hspace{-1.5mm}
             \end{array}
      \right]|\psi_{p}^{R}\rangle
      = \bold{0} .
\end{align}
Equation~(\ref{eq:v1=v2}) requires that the equations
\begin{align}
     \frac{\tilde{\eta}_{z}(\rho_{+})-E_{\bot}}
          {-i\tilde{\eta}_{y}(\rho_{+})}
 & = \frac{\tilde{\eta}_{z}(\rho_{-})-E_{\bot}}
          {-i\tilde{\eta}_{y}(\rho_{-})} ,
         \\
     \frac{-\tilde{\eta}_{z}(\rho_{+})-E_{\bot}}
          {-i\tilde{\eta}_{y}(\rho_{+})}
 & = \frac{-\tilde{\eta}_{z}(\rho_{-})-E_{\bot}}
          {-i\tilde{\eta}_{y}(\rho_{-})}
\end{align}
simultaneously hold.
This results in $E_{\bot} = 0$.

Setting $E_{\bot} = 0$ in Eq.~(\ref{eq:eve-appendix}),
we find that $\rho_{p}$ ($p = \pm$) is determined by
\begin{align}
  \tilde{M}-\frac{\rho_{p}+\rho_{p}^{-1}}{2}
  = \sigma \sqrt{1+\alpha^{2}}
    \left(\gamma-\frac{\rho_{p}-\rho_{p}^{-1}}{2}\right) ,
\end{align}
where $\sigma = \pm$ and
\begin{align}
    \label{eq:def-tilde-M}
  \tilde{M} = M -\frac{1}{2}(\beta_{x}+\beta_{x}^{-1}) .
\end{align}
Let us consider the case of $\sigma = -$.
Solving this equation with $\sigma = -$, we find that
\begin{align}
     \label{eq:rho-1}
  \rho_{1\pm}
  & = \frac{\sqrt{1+\alpha^{2}}\gamma+\tilde{M}
            \pm\sqrt{\left(\sqrt{1+\alpha^{2}}\gamma+\tilde{M}
                      \right)^{2}+\alpha^{2}}}
           {\sqrt{1+\alpha^{2}}+1}
\end{align}
and two eigenvectors satisfying
$|\psi_{+}^{R}\rangle = |\psi_{-}^{R}\rangle$ are given by
\begin{align}
     \label{eq:psi_B}
& |\psi_{{\rm B}1}^{R}\rangle
  = \xi \left[ \begin{array}{c}
                 \sqrt{1+\alpha^{2}} \\
                 1 \\
                 0 \\
                 \alpha
               \end{array}
        \right] ,
  \hspace{2mm}
  |\psi_{{\rm B}2}^{R}\rangle
  = \xi \left[ \begin{array}{c}
                 0 \\
                 -\alpha \\
                 \sqrt{1+\alpha^{2}} \\
                 1
               \end{array}
        \right] 
\end{align}
with $\xi = \left(2(1+\alpha^{2})\right)^{-\frac{1}{2}}$.
Substituting $\rho_{1\pm}$ and $|\psi_{{\rm B}q}^{R}\rangle$ ($q = 1$, $2$)
into Eq.~(\ref{eq:phi-R-localized}) and determining $c_{+}$ and $c_{-}$
in accordance with Eq.~(\ref{eq:condition1}), we obtain
\begin{align}
      \label{eq:Psi_1-B}
    |\Psi_{{\rm B}1}^{R}\rangle
  & = c \sum_{m}\beta_{x}^{m}
      \sum_{n=1}^{N}
      \left( \rho_{1+}^{n} - \rho_{1-}^{n} \right)
      |m,n \rangle \cdot |\psi_{{\rm B}1}^{R}\rangle ,
     \\
      \label{eq:Psi_2-B}
    |\Psi_{{\rm B}2}^{R}\rangle
  & = c \sum_{m}\beta_{x}^{m}
      \sum_{n=1}^{N}
      \left( \rho_{1+}^{n} - \rho_{1-}^{n} \right)
      |m,n \rangle \cdot |\psi_{{\rm B}2}^{R}\rangle ,
\end{align}
where $c$ is a normalization constant.
Let us turn to the case of $\sigma = +$.
Solving Eq.~(\ref{eq:def-tilde-M}) with $\sigma = +$, we find that
\begin{align}
     \label{eq:rho-2}
  \rho_{2\pm}
  & = \frac{\sqrt{1+\alpha^{2}}\gamma-\tilde{M}
            \pm\sqrt{\left(\sqrt{1+\alpha^{2}}\gamma-\tilde{M}
                      \right)^{2}+\alpha^{2}}}
           {\sqrt{1+\alpha^{2}}-1}
\end{align}
and two eigenvectors satisfying
$|\psi_{+}^{R}\rangle = |\psi_{-}^{R}\rangle$ are given by
\begin{align}
     \label{eq:psi_T}
&  |\psi_{{\rm T}1}^{R}\rangle
  = \xi \left[ \begin{array}{c}
                 \sqrt{1+\alpha^{2}} \\
                 -1 \\
                 0 \\
                 -\alpha
               \end{array}
        \right] ,
  \hspace{2mm}
  |\psi_{{\rm T}2}^{R}\rangle
  = \xi \left[ \begin{array}{c}
                 0 \\
                 \alpha \\
                 \sqrt{1+\alpha^{2}} \\
                 -1
               \end{array}
        \right] .
\end{align}
Substituting $\rho_{2\pm}$ and $|\psi_{{\rm T}q}^{R}\rangle$ ($q = 1$, $2$)
into Eq.~(\ref{eq:phi-R-localized}) and determining $c_{+}$ and $c_{-}$
in accordance with Eq.~(\ref{eq:condition2}), we obtain
\begin{align}
      \label{eq:Psi_1-T}
    |\Psi_{{\rm T}1}^{R}\rangle
  & = c' \sum_{m}\beta_{x}^{m}
      \sum_{n=1}^{N}
      \left( \frac{\rho_{2+}^{n}}{\rho_{2+}^{N+1}}
             - \frac{\rho_{2-}^{n}}{\rho_{2-}^{N+1}} \right)
      |m,n \rangle \cdot |\psi_{{\rm T}1}^{R}\rangle ,
     \\
      \label{eq:Psi_2-T}
    |\Psi_{{\rm T}2}^{R}\rangle
  & = c' \sum_{m}\beta_{x}^{m}
      \sum_{n=1}^{N}
      \left( \frac{\rho_{2+}^{n}}{\rho_{2+}^{N+1}}
             - \frac{\rho_{2-}^{n}}{\rho_{2-}^{N+1}} \right)
      |m,n \rangle \cdot |\psi_{{\rm T}2}^{R}\rangle ,
\end{align}
where $c'$ is a normalization constant.
$|\Psi_{{\rm B}q}^{R}\rangle$ ($q = 1$, $2$) is localized near
the bottom edge at $n = 1$ when $|\rho_{1\pm}| <1$
and $|\Psi_{{\rm T}q}^{R}\rangle$ ($q = 1$, $2$)
is localized near the top edge at $n = N$ when $|\rho_{2\pm}| > 1$.

Here, $|\Psi_{{\rm B}q}^{R}\rangle$ and $|\Psi_{{\rm T}q}^{R}\rangle$
($q = 1$, $2$) are eigenvectors of $H$.
Indeed, we easily find
\begin{align}
  h^{(1)}|\psi_{{\rm B}1}^{R}\rangle
  & = \sqrt{1+\alpha^{2}}\eta_{x}|\psi_{{\rm B}1}^{R}\rangle ,
    \\
  h^{(1)}|\psi_{{\rm B}2}^{R}\rangle
  & = - \sqrt{1+\alpha^{2}}\eta_{x}|\psi_{{\rm B}2}^{R}\rangle ,
    \\
  h^{(1)}|\psi_{{\rm T}1}^{R}\rangle
  & = - \sqrt{1+\alpha^{2}}\eta_{x}|\psi_{{\rm T}1}^{R}\rangle ,
    \\
  h^{(1)}|\psi_{{\rm T}2}^{R}\rangle
  & = \sqrt{1+\alpha^{2}}\eta_{x}|\psi_{{\rm T}2}^{R}\rangle ,
\end{align}
showing that they are eigenstates of $H$ and
their energy dispersion relations are 
\begin{align}
      \label{eq:disp-B1}
 E_{{\rm B}1}(k_{x})
   & = \sqrt{1+\alpha^{2}}\eta_{x}(k_{x}) ,
        \\
      \label{eq:disp-B2}
 E_{{\rm B}2}(k_{x})
   & = - \sqrt{1+\alpha^{2}}\eta_{x}(k_{x}) ,
        \\
      \label{eq:disp-T1}
 E_{{\rm T}1}({k^{\prime}}_{\!\!x})
   & = - \sqrt{1+\alpha^{2}}\eta_{x}({k^{\prime}}_{\!\!x}) ,
        \\
      \label{eq:disp-T2}
 E_{{\rm T}2}({k^{\prime}}_{\!\!x})
   & = \sqrt{1+\alpha^{2}}\eta_{x}({k^{\prime}}_{\!\!x}) .
\end{align}

We next consider the system  [see Fig.~A-1(b)]
under the obc in the $x$-direction,
\begin{align}
  \varphi^{R}(0,n) & = \varphi^{R}(N+1,n) = 0 ,
\end{align}
and the mpbc in the $y$-direction,
\begin{align}
  \varphi^{R}(m,n+N) & = b^{N} \varphi^{R}(m,n) .
\end{align}
The four eigenstates are obtained similarly to the derivation of
Eqs.~(\ref{eq:Psi_1-B}), (\ref{eq:Psi_2-B}), (\ref{eq:Psi_1-T}),
and (\ref{eq:Psi_2-T}):
\begin{align}
      \label{eq:Psi_1-L}
    |\Psi_{{\rm L}1}^{R}\rangle
  & = c \sum_{n}\beta_{y}^{n}
      \sum_{m=1}^{N}
      \left( \rho_{1+}^{m} - \rho_{1-}^{m} \right)
      |m,n \rangle \cdot |\psi_{{\rm L}1}^{R}\rangle ,
     \\
      \label{eq:Psi_2-L}
    |\Psi_{{\rm L}2}^{R}\rangle
  & = c \sum_{n}\beta_{y}^{n}
      \sum_{m=1}^{N}
      \left( \rho_{1+}^{m} - \rho_{1-}^{m} \right)
      |m,n \rangle \cdot |\psi_{{\rm L}2}^{R}\rangle ,
     \\
      \label{eq:Psi_1-R}
    |\Psi_{{\rm R}1}^{R}\rangle
  & = c' \sum_{n}\beta_{y}^{n}
      \sum_{m=1}^{N}
      \left( \frac{\rho_{2+}^{m}}{\rho_{2+}^{N+1}}
             - \frac{\rho_{2-}^{m}}{\rho_{2-}^{N+1}} \right)
      |m,n \rangle \cdot |\psi_{{\rm R}1}^{R}\rangle ,
     \\
      \label{eq:Psi_2-R}
    |\Psi_{{\rm R}2}^{R}\rangle
  & = c' \sum_{n}\beta_{y}^{n}
      \sum_{m=1}^{N}
      \left( \frac{\rho_{2+}^{m}}{\rho_{2+}^{N+1}}
             - \frac{\rho_{2-}^{m}}{\rho_{2-}^{N+1}} \right)
      |m,n \rangle \cdot |\psi_{{\rm R}2}^{R}\rangle ,
\end{align}
where
\begin{align}
     \label{eq:psi_L}
& |\psi_{{\rm L}1}^{R}\rangle
  = \xi \left[ \begin{array}{c}
                 \sqrt{1+\alpha^{2}} \\
                 -i \\
                 0 \\
                 -i\alpha
               \end{array}
        \right] ,
  \hspace{2mm}
  |\psi_{{\rm L}2}^{R}\rangle
  = \xi \left[ \begin{array}{c}
                 0 \\
                 -i\alpha \\
                 \sqrt{1+\alpha^{2}} \\
                 i
               \end{array}
        \right] ,
     \\
     \label{eq:psi_R}
&  |\psi_{{\rm R}1}^{R}\rangle
  = \xi \left[ \begin{array}{c}
                 \sqrt{1+\alpha^{2}} \\
                 i \\
                 0 \\
                 i\alpha
               \end{array}
        \right] ,
  \hspace{2mm}
  |\psi_{{\rm R}2}^{R}\rangle
  = \xi \left[ \begin{array}{c}
                 0 \\
                 i\alpha \\
                 \sqrt{1+\alpha^{2}} \\
                 -i
               \end{array}
        \right] .
\end{align}
$|\Psi_{{\rm L}q}^{R}\rangle$ ($q = 1$, $2$) is localized near
the left edge at $m = 1$ when $|\rho_{1\pm}| <1$
and $|\Psi_{{\rm R}q}^{R}\rangle$ ($q = 1$, $2$)
is localized near the right edge at $m = N$ when $|\rho_{2\pm}| > 1$.
Their energy dispersion relations are
\begin{align}
      \label{eq:disp-L1}
 E_{{\rm L}1}(k_{y})
   & = - \sqrt{1+\alpha^{2}}\eta_{y}(k_{y}) ,
        \\
      \label{eq:disp-L2}
 E_{{\rm L}2}(k_{y})
   & = \sqrt{1+\alpha^{2}}\eta_{y}(k_{y}) ,
        \\
      \label{eq:disp-R1}
 E_{{\rm R}1}({k^{\prime}}_{\!\!y})
   & = \sqrt{1+\alpha^{2}}\eta_{y}({k^{\prime}}_{\!\!y}) ,
        \\
      \label{eq:disp-R2}
 E_{{\rm R}2}({k^{\prime}}_{\!\!y})
   & = - \sqrt{1+\alpha^{2}}\eta_{y}({k^{\prime}}_{\!\!y}) .
\end{align}

Let us consider helical edge states in the boundary geometry.
A pair of helical edge states circulate
along the square loop consisting of the bottom, top, left, and right edges.
Helical edge states circulating in the anti-clockwise direction
are allowed to form a subband only when the four dispersion relations
Eqs.~(\ref{eq:disp-B1}), (\ref{eq:disp-T1}), (\ref{eq:disp-L1}), and
(\ref{eq:disp-R1}) become identical for the appropriately
chosen $k_{x}$, ${k^{\prime}}_{\!\!x}$, $k_{y}$, and ${k^{\prime}}_{\!\!y}$
[see Fig.~A-1(c)]:
\begin{align}
  E_{{\rm B}1}(k_{x}) = E_{{\rm T}1}({k^{\prime}}_{\!\!x})
  = E_{{\rm L}1}(k_{y})= E_{{\rm R}1}({k^{\prime}}_{\!\!y}) .
\end{align}
Considering the expressions of $\eta_{x}$ and $\eta_{y}$ given in
Eqs.~(\ref{eq:def-eta_x-mod}) and (\ref{eq:def-eta_y-mod}), respectively,
we find that this holds when
\begin{align}
    \label{eq:cond-helical1}
 k_{x} = -{k^{\prime}}_{\!\!x} = -k_{y} = {k^{\prime}}_{\!\!y} \equiv k
\end{align}
with
\begin{align}
    \label{eq:cond-helical2}
  b_{-}\cos(ka) = \gamma .
\end{align}
Equation~(\ref{eq:cond-helical2}) with Eqs.~(\ref{eq:def-eta_x-mod})
and (\ref{eq:def-eta_y-mod}) requires that
the energy of the helical edge states must be real.
The resulting dispersion relation is given by
\begin{align}
  E_{1}(k) = \sqrt{1+\alpha^{2}}
             \sqrt{1+\left(\frac{\gamma}{\cos(ka)}\right)^{2}}
             \sin(ka) .
\end{align}
The other helical edge states circulating in the clockwise direction
are allowed to form a subband only when the four dispersion relations
Eqs.~(\ref{eq:disp-B2}), (\ref{eq:disp-T2}), (\ref{eq:disp-L2}), and
(\ref{eq:disp-R2}) become identical for the appropriately
chosen $k_{x}$, ${k^{\prime}}_{\!\!x}$, $k_{y}$, and ${k^{\prime}}_{\!\!y}$
[see Fig.~A-1(d)]:
\begin{align}
  E_{{\rm B}2}(k_{x}) = E_{{\rm T}2}({k^{\prime}}_{\!\!x})
  = E_{{\rm L}2}(k_{y}) = E_{{\rm R}2}({k^{\prime}}_{\!\!y}) .
\end{align}
This again holds under the condition given in
Eqs.~(\ref{eq:cond-helical1}) and (\ref{eq:cond-helical2}).
The resulting dispersion relation is given by
\begin{align}
  E_{2}(k) = - \sqrt{1+\alpha^{2}}
             \sqrt{1+\left(\frac{\gamma}{\cos(ka)}\right)^{2}}
             \sin(ka) .
\end{align}

We find that the energy of a helical edge state must be real
although $\gamma \neq 0$.
This is the most important characteristic of the helical edge states
in our system.
Another important characteristic is derived from Eq.~(\ref{eq:cond-helical2})
that the wavefunction amplitude varies
in the $x$- and $y$-directions with the same rate of increase:
\begin{align}
     \label{eq:b-CEs_appendix}
  b = \sqrt{1+\left(\frac{\gamma}{\cos (ka)}\right)^{2}}
      + \frac{\gamma}{\cos (ka)} .
\end{align}
These two characteristics are used in Sect.~4.

In the systems shown in Figs.~A-1(a) and A-1(b), the helical edge states
are stabilized when $|\rho_{1\pm}| <1$ and $|\rho_{2\pm}| >1$.
However, we do not need to take into account these conditions
when considering the helical edge states in the boundary geometry
because they are less strict than the condition requiring that
the energy of each helical edge state is real.

\end{document}